\documentclass{svjour3}

\let\vec\relax 
\DeclareMathAccent{\vec}{\mathord}{letters}{"7E} 

\usepackage[utf8]{inputenc}
\usepackage[english]{babel}
\usepackage{amsmath}
\usepackage{amssymb}
\usepackage{mathtools}
\usepackage{algorithm}
\usepackage{algorithmic}
\usepackage[binary-units=true]{siunitx}
\usepackage{enumerate}
\usepackage{soul}

\usepackage[square,numbers]{natbib}
\usepackage{hyperref}
\hypersetup{colorlinks=true,urlcolor=blue,citecolor=blue,linkcolor=blue,breaklinks=true}

\usepackage{fullpage}
\usepackage{booktabs}
\usepackage{array}

\usepackage{tikz}
\usepackage{xcolor}
\usetikzlibrary{patterns}
\usetikzlibrary{arrows,decorations.pathreplacing}
\usepackage{pgfplots}
\usepackage{url}
\usetikzlibrary{decorations.pathmorphing}
\tikzset{every picture/.style=remember picture}

\newcommand{\abs}[1]{|#1|}
\newcommand{\R}{\mathbb{R}}
\newcommand{\N}{\mathbb{N}}
\newcommand{\C}{\mathbb{C}}
\newcommand{\norm}[2]{\left\|#1\right\|_{#2}}
\renewcommand{\d}[1]{\textrm{ d}#1}
\newcommand{\set}[2]{\left\{#1~|~#2\right\}}
\renewcommand{\Re}[1]{\textrm{Re}(#1)}
\renewcommand{\Im}[1]{\textrm{Im}(#1)}

\newcommand{\iu}{\mathrm{i}\mkern1mu}
\newcommand{\bra}[1]{\langle#1|}
\newcommand{\ket}[1]{|#1\rangle}

\newcommand{\operator}[1]{\hat{#1}}
\newcommand{\creation}[1]{\operator{c}^\dagger_{#1}}
\newcommand{\annihilation}[1]{\operator{c}^{\phantom\dagger}_{#1}}
\newcommand{\occupation}[1]{\operator{n}_{#1}}
\newcommand{\doubleoccupation}[1]{\operator{d}_{#1}}
\newcommand{\spinup}{\uparrow}
\newcommand{\spindown}{\downarrow}
\newcommand{\nsites}{{N_s}}
\newcommand{\nup}{{n_\spinup}}
\newcommand{\ndown}{{n_\spindown}}
\newcommand{\nstates}{{N_\psi}}
\newcommand{\hilbertspace}{\mathcal{H}}
\newcommand{\numsci}[2]{$#1\,\scriptstyle{\textrm{E}#2}$}
\newcommand{\mathnode}[2]{\mathord{\tikz[baseline=(#1.base), inner sep = 0pt]{\node (#1) {$#2$};}}}
\newcommand{\row}{\mathcal{I}}
\newcommand{\col}{\mathcal{J}}
\newcommand{\val}{\mathcal{V}}
\newcommand{\g}[1]{\textcolor{gray}{#1}}
\newcommand{\gl}[1]{\textcolor{gray!50}{#1}}
\newcommand{\groundstate}{\vec{v}_0}
\newcommand{\tol}{\mathrm{tol}}
\newcommand*\colvec[3][]{
    \begin{pmatrix}\ifx\relax#1\relax\else#1\\\fi#2\\#3\end{pmatrix}
}

\begin{document}

\title{Electron-light interaction in nonequilibrium -- exact diagonalization for time dependent Hubbard Hamiltonians}
\author{Michael Innerberger \and Paul Worm \and  Paul Prauhart \and Anna Kauch}

\institute{M. Innerberger \at Institute of Analysis and Scientific Computing, Vienna University of Technology, Wiedner Hauptstr. 8-10, 1040 Wien, Austria \and
P. Worm \at
Institute of Solid State Physics, Vienna University of Technology, Wiedner Hauptstr. 8-10, 1040 Wien, Austria \and 
P. Prauhart \at  
Institute of Solid State Physics, Vienna University of Technology, Wiedner Hauptstr. 8-10, 1040 Wien, Austria \and
A. Kauch \at
Institute of Solid State Physics, Vienna University of Technology, Wiedner Hauptstr. 8-10, 1040 Wien, Austria \\
\email{kauch@ifp.tuwien.ac.at}
}

\date{\today}

\maketitle

\abstract{
We present a straightforward implementation scheme for solving the time dependent Schr\"odinger equation for systems described by the Hubbard Hamiltonian with time dependent hoppings. The computations can be performed for clusters of up to $14$ sites with in principle general geometry. For the time evolution, we use the exponential midpoint rule, where the exponentials are computed via a Krylov subspace method, which only uses matrix-vector multiplication. The presented implementation uses standard libraries for constructing sparse matrices and for linear algebra. Therefore, the approach is easy to use on both desktop computers and computational clusters. We apply the method to calculate time evolution of double occupation and nonequilibrium spectral function of a photo-excited Mott-insulator. The results show that not only the double occupation  increases due to creation of electron-hole pairs but also the Mott gap becomes partially filled.
}

\section{Introduction}\label{ch:intro}

Photo-induced states of matter gain increasing attention for their exotic properties~\cite{exp1,exp2,exp3,Kaneko_eta_pairing,Fehske,Wang2018} and possible applications, e.g. in the context of energy conversion~\cite{impact2,impact}. The description of these states necessitates nonequilibrium approaches, which are particularly demanding in cases where light brings a strongly correlated electronic system out of equilibrium. The approximate theoretical approaches to correlated systems are being successfully adapted to treat systems out of equilibrium (e.g. nonequilibrium dynamical mean-field theory (DMFT)~\cite{dmft}, dynamical cluster approximation~\cite{dca}, auxiliary master equation approach~\cite{arrigoni}, GW~\cite{GW}). The numerically exact approaches, exact diagonalization (ED)~\cite{ED}, or density-matrix renormalization group~\cite{White,Fehske}, where the error can be systematically controlled, are still limited to relatively small system sizes or short times~\cite{dmrg_review1}. They are, however, invaluable for benchmarking sophisticated approximate methods. The purpose of this paper is to present a straightforward implementation of the ED method using well-known data formats and algorithms in order to employ highly-optimized libraries. The method currently allows for calculations with up to $14$ sites. 

We specifically focus on the application of the method to calculate electronic properties of a system that is described by a time-dependent Hubbard Hamiltonian. The time dependence is introduced by coupling of the electronic system to a electromagnetic (EM) field pulse. The EM field is treated classically and enters the hoppings via Peierls' substitution. We apply the method to study the  time evolution of a Mott-insulator after interaction with a light-pulse. By calculating double occupation and the nonequilibrium spectral function, we show photo-doping of the original Mott-insulator~\cite{photo_doping,Eckstein2016} as well as filling of the Mott gap~\cite{impact,quantum_boltzmann,photo_doping_aw,Golez2015}. 

The paper is organized as follows. In Sec.~\ref{ch:model} we introduce the Hamiltonian of the Hubbard model with Peierls' substitution, the observables of our interest, as well as notation and units. In Sec.~\ref{ch:implementation} we give a detailed description of data formats and the time stepping algorithm as well as how observables are practically calculated. In Sec.~\ref{ch:results} we present the time evolution of double occupation and nonequilibrium spectral function to illustrate the application of the method. In Sec.~\ref{ch:outlook} we give a short summary and outlook. 



\section{Model}\label{ch:model}
\subsection{Hubbard model}
We focus on the paradigm model for strongly correlated electrons, the Hubbard model~\cite{hubbard}, given by the following Hamiltonian:
\begin{equation}
\label{eq:hubbardhamiltonian}
\operator{H}_\textrm{Hubbard} = 
 -\sum_{i,j,\sigma} v_{ji}\creation{i\sigma}\annihilation{j\sigma} +
 U \sum_i \occupation{i\spinup}\occupation{i\spindown},
\end{equation}
where 
$v_{ji}$ describes the relative probability amplitude of an electron hopping from site $j$ to $i$ without change of spin; $U>0$ is the on-site Coulomb repulsion between two electrons if they reside at the same site (with opposite spins); $\creation{i\sigma}$ ($\annihilation{i\sigma}$) denote the fermionic  creation (annihilation) operators at site $i$ with spin $\sigma$ and $\occupation{i\sigma}=\creation{i\sigma} \annihilation{i\sigma}$ is the occupation number operator (for details on second quantization formalism c.f. \cite{manybody}). 

\begin{figure}
	\centering
	\includegraphics[width=0.6\linewidth]{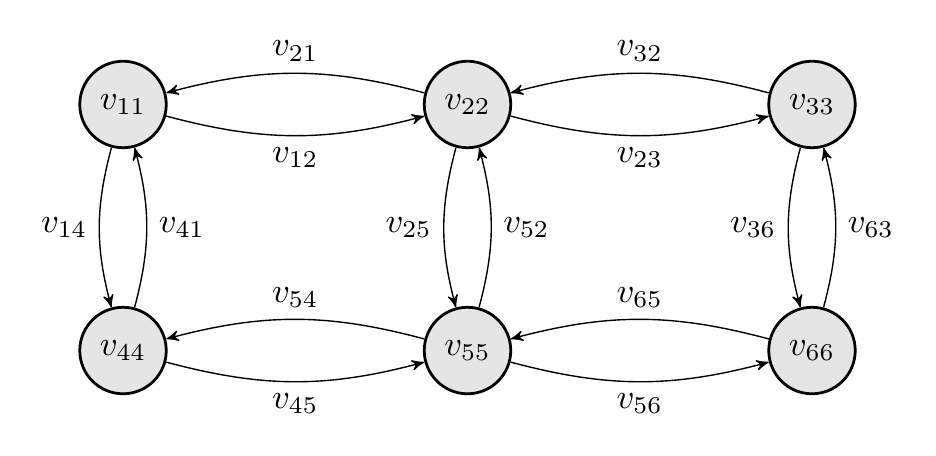}
	\caption{Example geometry of a two-dimensional six-site lattice with lexicographical ordering of the sites. The energies $v_{ii}$ describe an additional on-site potential, $v_{ij}$ describe the hoppings between sites $i$ and $j$.}
	\label{fig:geometry}
\end{figure}

In the following, we restrict our considerations to finite size systems of $\nsites\in \N$ sites with hoppings explicitly given by a hopping matrix $v = (v_{ij})_{i,j=1}^\nsites$. The hopping matrix can be arbitrary, i.e. we can allow for finite hoppings between any two sites. This is where the geometry of the studied system is encoded and where also periodic boundary conditions can be introduced. Lattices of arbitrary dimension and shape can be studied with this approach. Additionally, we can introduce on-site potentials, which can be added as diagonal elements $v_{ii}$ of the hopping matrix. Fig.~\ref{fig:geometry} illustrates a  $2\times 3$  box geometry with open boundary conditions.  

\subsection{Time dependent electron-light interaction}

The interaction of electrons with light puts the system out of equilibrium. Here, the light is modeled as a classical electric field pulse
\begin{equation}\label{eq:efield}
	\vec{E}(t)=\vec{E}_0\sin(\omega(t-t_p))e^{-\frac{(t-t_p)^2}{2 \sigma^2}}
\end{equation}
of width $\sigma$, peaked around the time $t_p$, and with frequency $\omega$. We set the units of frequency equal to the units of energy ($\hbar \equiv 1$) and the unit of time is then the inverse of the unit of energy. The EM field is included in the Hubbard Hamiltonian using Peierls' substitution~\cite{Peierls}, which adds a time dependence to the hoppings:
\begin{equation} \label{eq:peierls}
	v_{ij} \rightarrow v_{ij}(t)=v_{ij}\exp\left({-\iu e \int_{\vec{R}_i}^{\vec{R}_j}\vec{A}(\vec{r}',t)d\vec{r}'}\right).
\end{equation}
We use a gauge where the scalar potential vanishes and $\vec{E}=\partial_t\vec{A}(t)$. In general, the result of the integral in Eq.~\eqref{eq:peierls} will depend on the direction of the $\vec{E}$-field and on the relative position of sites $i$ and $j$. The time-dependent phase factor must then be defined for each pair of sites $i$ and $j$ separately. 

In the simpler case of only nearest neighbour (NN) hopping and box geometry, the integral will only depend on whether the hopping between NN sites $i$ and $j$ is in the horizontal or vertical direction. By choosing the $\vec{E}$-field direction to be diagonal with respect to the box, we can describe the time dependence  by only one function $f(t)$ for each non-zero element of the hopping matrix $v$.
For the sake of simplicity we further approximate the integral in Eq.~\eqref{eq:peierls} to arrive at
\begin{equation} \label{eq:vt}
	f(t) =  \exp\left(\iu a\left[\cos(\omega(t-t_p))-b\right]e^{-\frac{(t-t_p)^2}{2\sigma^2}}\right)
\end{equation}
%
with dimensionless parameters $a$ and $b$. The parameter $a$ describes the strength of the EM field, whereas $b$ can be used to set the initial phase factor of the hoppings to $1$. Please note, that the Peierls' substitution introduces only a phase factor to the hoppings and does not change their absolute value. For all the results presented the NN hoppings will be set to have equal absolute value and this value is used as the unit of energy, i.e.\ $|v_{ij}| = 1$.

%
%

\subsection{Symmetries of the Hamiltonian}\label{sec:symmetries}


Allowing for at most two electrons (with different spins) per site, a state of the system can be represented by the state vector $\ket{\psi} = \ket{n_{1\spinup} n_{1\spindown} n_{2\spinup} \ldots n_{\nsites\spindown}}$, where $n_{i\sigma} \in \{0,1\}$ is the number of electrons with spin $\sigma$ at site $i$.
All states of this form are orthonormal and form an abstract Hilbert space which we denote by $\hilbertspace(\nsites)$.
The subspace of all states with $\nup$ electrons with spin up and $\ndown$ electrons with spin down is denoted by $\hilbertspace^\nup_\ndown(\nsites)$.
It is easy to see that there holds (with $\oplus$ being the direct sum)
\begin{equation}\label{eq:directsum}
\hilbertspace(\nsites) = \bigoplus_{0\leq \nup, \ndown \leq \nsites} \hilbertspace^\nup_\ndown(\nsites).
\end{equation}

Any state in $\hilbertspace(\nsites)$ can be seen as excitation of the vacuum state:
\begin{equation}\label{eq:statedef}
(\creation{1\spinup})^{n_{1\spinup}} \cdot
(\creation{1\spindown})^{n_{1\spindown}} \cdot
\ldots \cdot
(\creation{\nsites\spindown})^{n_{\nsites\spindown}}
\ket{00\ldots 0} = 
\ket{n_{1\spinup} n_{1\spindown} \ldots n_{\nsites\spindown}},
\end{equation}
where the action of fermionic creation and annihilation operators $\creation{}$ and $\annihilation{}$ on a particular state is given by
\begin{alignat}{2}
\label{eq:creation-annihilation}
\creation{i\sigma} \ket{n_{1\spinup}  \ldots n_{i\sigma} \ldots n_{\nsites\spindown}}
&=&
~\delta_{n_{i\sigma},0} (-1)^{c_{i\sigma}} \sqrt{n_{i\sigma}+1} ~
&\ket{n_{1\spinup}  \ldots (n_{i\sigma}+1) \ldots n_{\nsites\spindown}},\\
\annihilation{i\sigma} \ket{n_{1\spinup}  \ldots n_{i\sigma} \ldots n_{\nsites\spindown}}
&=&
~\delta_{n_{i\sigma},1} (-1)^{c_{i\sigma}} \sqrt{n_{i\sigma}} ~
&\ket{n_{1\spinup}  \ldots (n_{i\sigma}-1) \ldots n_{\nsites\spindown}},
\end{alignat}
respectively.
Here, $c_{i\sigma}$ is the number of how many electrons are present in the state up to the $i\sigma$-entry.
This is due to the fermionic anticommutator relations, 
which cause that switching the order of two adjacent operators results in an additional negative sign.
Note that the definition~\eqref{eq:statedef} is not consistent throughout literature.
Finally, the action of the number operator $\occupation{i\sigma} = \creation{i\sigma} \annihilation{i\sigma}$ is given by the equation
\begin{equation}
\occupation{i\sigma} \ket{n_{1\spinup} n_{1\spindown} \ldots n_{\nsites\spindown}} = n_{i\sigma} \ket{n_{1\spinup} n_{1\spindown} \ldots n_{\nsites\spindown}},
\end{equation}
i.e.\ this operator counts  the number of electrons  on site $i$ with spin $\sigma$.

Because the Hubbard Hamiltonian commutes with the occupation number and spin operators, the number of electrons of spin $\sigma$ in the system $\sum_{i} \occupation{i\sigma}$ is invariant under the Hamiltonian in Eq.~\eqref{eq:hubbardhamiltonian}.
This means that, in the basis of all states in the Hilbert space $\hilbertspace(\nsites)$, the Hamiltonian takes a block-diagonal form, according to the direct sum in Eq.~\eqref{eq:directsum}. Our implementation exploits this block-diagonal form and generates the Hamiltonian only in the requested subspace $\hilbertspace^\nup_\ndown(\nsites)$, with arbitrary $\nsites,\nup$ and $\ndown$. Since we are interested in Mott-insulators, we take the system to be half-filled, i.e.\ $\nup=\ndown=\nsites/2$ and the total spin is zero.

\subsection{Time evolution and observables}

The Hubbard Hamiltonian with time-dependent hoppings is a time-dependent Hermitian operator that describes the evolution of a state $\ket{\psi_0} \in \hilbertspace(\nsites)$ in terms of the Schr\"odinger equation ($\hbar\equiv 1$)
\begin{equation}
\label{eq:schroedinger}
\iu \partial_t \ket{\psi(t)} = \operator{H}(t) \ket{\psi(t)},
\qquad \ket{\psi(0)}=\ket{\psi_0}.
\end{equation}
Exact diagonalization means that Eq.~\eqref{eq:schroedinger} is solved over the finite dimensional Hilbert space $\hilbertspace(\nsites)$, which yields a large system of ordinary differential equations.
The exact solution is given by
\begin{equation}
\label{eq:exact-solution}
	\ket{\psi(t)}
	= \mathcal{T} e^{-\iu \int_0^t \operator{H}(\tau) \d{\tau}} \ket{\psi_0},
\end{equation}
where $\mathcal{T}$ is the time ordering operator \cite{manybody}.
Once  the state of the system at time $t$, $\ket{\psi(t)}$, is known the expectation value of an observable $\hat{O}$ can be  calculated directly through
\begin{equation}\label{eq:observable}
\langle\hat{O}(t)\rangle = \bra{\psi(t)}\hat{O}\ket{\psi(t)}.
\end{equation}

Specifically, we are interested in the (time-dependent) average double occupation per site
\begin{equation}\label{eq:docc}
\langle\doubleoccupation{}(t)\rangle = \tfrac{1}{\nsites} \sum_{i=1}^{\nsites}\bra{\psi(t)}\doubleoccupation{i}\ket{\psi(t)},
\end{equation}
with $\doubleoccupation{i}=\occupation{i\spinup}\occupation{i\spindown}$, and the average energy per site  
\begin{equation}\label{eq:energy}
E(t)=\langle\operator{H}(t)\rangle = \tfrac{1}{\nsites} \bra{\psi(t)}\operator{H}(t)\ket{\psi(t)}. 
\end{equation}
In the following, we drop the explicit time dependencies if it is clear from context.
To numerically obtain these quantities of interest, one must assemble a matrix representation of $\operator{H}$, compute the ground state and then carry out a time stepping before building the expectation values.
These tasks are computationally not trivial, since the number of independent variables grows exponentially in the number of sites $\nsites$.
Note, however, that one can still treat the subspaces $\hilbertspace_\ndown^\nup(\nsites)$ separately, because the time-dependent Hamiltonian $\operator{H}(t)$ commutes with $\sum_i \occupation{i\sigma}$ (i.e.\ it preserves the number of electrons).

\subsection{Nonequilibrium spectral function}

The time-stepping algorithm allows also for calculation of double-time correlation functions. For example the nonequilibrium Green's functions $G^<$ and $G^>$ are obtained through~\cite{dmft} 
\begin{equation}
    \begin{split}
        &G_{ij\sigma}^{<}(t,t') = \iu  \bra{\psi(t')}\creation{j\sigma} \mathcal{T} e^{-\iu\int_{t}^{t'} H(\tau) \d{\tau}} \;\annihilation{i\sigma}  \ket{\psi(t)} ,\\
        &G_{ij\sigma}^{>}(t,t') = -\iu  \bra{\psi(t)} \annihilation{i\sigma} \mathcal{T} e^{-\iu\int_{t'}^{t} H(\tau) \d{\tau}} \;\creation{j\sigma} \ket{\psi(t') },\\
    \end{split}
\label{eq:Glg}
\end{equation}
where $\mathcal{T}$ is the time ordering operator and $\ket{\psi(t)}$ is the solution of Eq.~\eqref{eq:schroedinger}. In order to obtain the correlation functions in Eq.~\eqref{eq:Glg}, we must apply the annihilation (creation) operator $\annihilation{i\sigma}\;$  ($\creation{j\sigma}$) to $\ket{\psi(t)}$ and then time-evolve the resulting state according to Eq.~\eqref{eq:schroedinger} in the subspace with one electron less (more), act again with  $\creation{j\sigma}\;$ ($\annihilation{i\sigma}$) on the result, and finally build the expectation value.

The nonequilibrium spectral function \cite{dmft} $A(\nu,t) =  \big( A^{<}(\nu,t) + A^{>}(\nu,t) \big)$ can then be obtained after performing a forward Fourier transform of $G^{\stackrel{<}{>}}(t,t')$~\cite{photo_doping_aw}: 

\begin{equation}
    A^{\stackrel{<}{>}}(\nu,t) =  \frac{1}{\pi} \text{Im} \int_0^{\infty} e^{\iu \nu t_{\text{rel}}} G^{\stackrel{<}{>}}(t,t') \d{t_{\text{rel}}}
\label{eq:A_noneq}
\end{equation}
with  $t_{\text{rel}} = t' - t$ (we omitted spin and site indices for simplicity).

\paragraph{Lehmann representation}

In equilibrium, the spectral function must remain time-independent and can be benchmarked against the Lehmann representation. The site-averaged spectral function is then given by
\begin{equation}\label{eq:spectralfunction}
A(\nu) = \frac{1}{\nsites}\sum_{i,\sigma} \sum_{\ket{\phi}}
\left(|\bra{\phi}\creation{i\sigma}\ket{\psi_0}|^2\delta(\nu-E_{\ket{\phi}}+E_0)
+|\bra{\phi}\annihilation{i\sigma}\ket{\psi_0}|^2\delta(\nu+E_{\ket{\phi}}-E_0)\right)
\end{equation}
Here, $\{\ket{\phi}\}$ is an eigenbasis of $\hilbertspace(\nsites)$ with respective energy eigenvalues $E_{\ket{\phi}}$, and $\ket{\psi_0}$ is the ground state of $\operator{H}$ with energy $E_0$.

\section{Implementation}\label{ch:implementation}

The aim of the present paper is to provide efficient data structures and algorithms for assembling matrix representations of Hubbard-Hamiltonians for arbitrary problems of the type that was introduced above, as well as a simple time-stepping algorithm for solving the arising time-dependent Schr\"odinger equation, Eq.~\eqref{eq:schroedinger}.
In the following, the key points of the implementation are discussed.
For linear algebra subroutines, existing libraries such as Intel's MKL, and LAPACK / BLAS are used, as well as the matrix exponentiation library Expokit \cite{expokit}, which is an essential part of the time-stepping algorithm.

\subsection{Discrete basis of subspace $\hilbertspace^\nup_\ndown(\nsites)$}\label{sec:states}

\paragraph{Number of states}

Due to spin-up and spin-down electrons being independent, $\hilbertspace^\nup_\ndown(\nsites)$ can be identified with $\hilbertspace_{0}^{\nup}(\nsites) \otimes \hilbertspace_{\ndown}^{0}(\nsites)$.
The problem of how to place $\nup$ ($\ndown$) electrons on $\nsites$ sites is well known in combinatorics.
This leads to
\begin{equation}
\label{eq:statedim}
	\nstates(\nup,\ndown)
	= \dim \Big( \hilbertspace_{\ndown}^{\nup}(\nsites) \Big)
	= \dim\Big(\hilbertspace_{0}^{\nup}(\nsites)\Big) \dim\Big(\hilbertspace_{\ndown}^{0}(\nsites)\Big)
	= \binom{\nsites}{\nup}\binom{\nsites}{\ndown}.
\end{equation}
Note that $\nstates = \nstates(\nup,\ndown)$ takes a maximum for $\nup=\ndown=\nsites/2$.
For the number of all states there holds
\begin{equation}
\label{eq:number-of-states}
	\dim (\hilbertspace(\nsites)) = 2^{2\nsites} = 4^\nsites,
\end{equation}
since there are $2\nsites$ vacancies in the lattice that can be occupied or not.
This suggests that the general computational effort for assembling a Hamiltonian and for time-stepping on a system with $\nsites$ sites scales at least like $\mathcal{O}(4^\nsites)$. For different $\nsites$, the value of $\nstates$ is shown in Table~\ref{tab:dimensions}.

\paragraph{State representation}

On a computer, the states can be represented by $2\nsites$ bits, which can be stored internally in integers of sufficient size.
All actions like hopping, creation, and annihilation of electrons can then be implemented as bitwise operations. 

To obtain all states that constitute a basis $\mathcal{B}$ of $\hilbertspace_\ndown^\nup(\nsites)$ for fixed numbers $\nup$, $\ndown$, and $\nsites$, the hopping of electrons is emulated.
Starting from an initial state with the right number of electrons, all other states can be obtained by repeated hopping (i.e.\ flipping two bits). 
Due to the independence of spin-up and spin-down electrons, we can treat $\hilbertspace_\ndown^0(\nsites)$ and $\hilbertspace_0^\nup(\nsites)$ separately and get all states from building the tensor-product of the respective bases $\mathcal{B}_\spinup$ and $\mathcal{B}_\spindown$.
Therefore, we restrict the presentation to the case of spin-up electrons in the following.

A multi-index $\alpha \in \{1,\ldots,\nsites\}^{\nup}$ can be used to represent the (ordered) positions of electrons on the sites, i.e.\ $\alpha_i = j$ means that the $i$-th electron resides at site $j$.
From Pauli's principle we see that
\begin{equation}
\label{eq:multilimits}
	1 \leq \alpha_1 < \ldots < \alpha_\nup \leq \nsites.
\end{equation}
For such multi-indices one can define a total ordering by
\begin{equation}
\label{eq:multi-index-ordering}
	\alpha < (>)~\hat{\alpha}
	\quad \Longleftrightarrow \quad
	\alpha_j < (>)~\hat{\alpha}_j
	\quad\text{with }
	j = \arg\min \set{\alpha_i \neq \hat{\alpha}_i}{i = 1,\ldots,\nup}.
\end{equation}
This gives a natural meaning to \emph{increasing $\alpha$ by one}.
From Eq.~\eqref{eq:multilimits} we see further that the smallest admissible multi-index satisfies $\alpha_i = i$ and the largest satisfies $\alpha_i = \nsites-\nup-i$ for all $i = 1,\ldots,\nup$.
By iterating over all multi-indices yielding to the limitation posed by Eq.~\eqref{eq:multilimits}, one obtains all possible permutations of electrons.
This is shown in pseudo-code in Alg.~\ref{alg:genstates}.

\begin{algorithm}[h]
	\caption{Generating states with only spin-up excitations}
	\label{alg:genstates}
	\textbf{Input:} $\nup$, $\nsites$\\
	\textbf{Output:} $\mathcal{B}_\spinup$
	\begin{algorithmic}[1]
		\STATE $\mathcal{B}_\spinup = \emptyset$
		\STATE $\alpha_i = i$ for $i = 1,\ldots,\nup$
		\STATE $\ket{\psi_\textrm{init}} = \ket{11\ldots0}$ ~/* ones up to the $\nup$-th position */
		\WHILE{$\alpha$ can be further increased}
			\STATE $\ket{\psi} = \ket{\psi_\textrm{init}}$
			\FOR{$i=1$ \TO $\nup$}
				\STATE /* hop from $i$-th to $\alpha_i$-th position in $\ket{\psi}$ */
				\STATE $\psi[i] = 0$, $\psi[\alpha_i] = 1$
			\ENDFOR
			\STATE $\mathcal{B}_\spinup = \mathcal{B}_\spinup \cup \ket{\psi}$
			\STATE increase $\alpha$ by one
		\ENDWHILE
	\end{algorithmic}
\end{algorithm}

\subsection{Sparse structure of the Hamiltonian}\label{sec:hamiltonian}

Because of Eq.~\eqref{eq:number-of-states}, even for small $\nsites$ the matrix representation of the Hamiltonian for most electron configurations requires vast amounts of memory if implemented as double-precision complex matrix.
Due to the limited overlap of states, many elements of the matrix representation are zero.
Utilizing this fact allows for using  a well-known sparse matrix format, resulting in a much more memory efficient implementation.

\paragraph{Non-zero elements}

Consider fixed numbers $\nsites$, $\nup$, and $\ndown$.
Conservation of electron number implies that two states in $\hilbertspace^{n_\spinup}_{n_\spindown}(\nsites)$ can only differ by an even number of entries.
Furthermore, hopping between more than two sites is not accounted for by the Hubbard Hamiltonian in Eq.~\eqref{eq:hubbardhamiltonian}.
This leaves only two cases that give a non-zero contribution.

First, a state differs from itself by zero entries, which gives a  contribution to the diagonal of the Hamiltonian.
Second, one of the $n_\spinup$ ($n_\spindown$) spin up (spin down) electrons can hop to one of the $\nsites - n_\spinup$ ($\nsites - n_\spindown$) unoccupied sites, creating a state differing in exactly two entries.
There are $n_\spinup(\nsites-n_\spinup)$ ($n_\spindown(\nsites-n_\spindown)$) possibilities for that process, each giving an off-diagonal non-zero contribution to the Hamiltonian.
Due to the hermiticity of the Hamiltonian, only its upper triangular part, which consists of the diagonal and half of all off-diagonal non-zero elements, yields non-redundant information. Thus, the number of non-zero elements evaluates to
\begin{equation}
\label{eq:non-zero-entries}
	N_\textrm{nz} = \nstates(1 + \tfrac{1}{2}\nup(\nsites-\nup) + \tfrac{1}{2}\ndown(\nsites-\ndown)).
\end{equation}

Note that Eq.~\eqref{eq:non-zero-entries} is only a worst-case result. If some of the coefficients $U$, $v_{ij}$, or combinations thereof are zero, this further reduces the number of nontrivial entries of the Hamiltonian's matrix representation. For the case of half-filling, we get $N_\textrm{nz} = \nstates(1+\tfrac{\nsites^2}{4})$. Together with $\nstates = \mathcal{O}(4^\nsites)$, this means that the non-zero elements can be stored with $\mathcal{O}(\nstates\ln^2(\nstates))$ memory, which is nearly linear, as opposed to quadratic memory $\mathcal{O}(\nstates^2)$ for dense matrices.

\begin{table}
	\renewcommand{\arraystretch}{1.3}
	\begin{centering}
		\begin{tabular}{l|cccccccc}
			$\nsites$ & 2 & 4 & 6 & 8 & 10 & 12 & 14 & 16 \\ 
			\hline\hline
			$\nstates$ & \numsci{4.0}{+0} & \numsci{3.6}{+1} & \numsci{4.0}{+2} & \numsci{4.9}{+3} & \numsci{6.4}{+4} & \numsci{8.5}{+5} & \numsci{1.2}{+7} &  \numsci{1.7}{+8}\\ 
			$N_{\textrm{nz}}$ & \numsci{8.0}{+0} & \numsci{1.8}{+2} & \numsci{4.0}{+3} & \numsci{8.3}{+4} & \numsci{1.7}{+6} & \numsci{3.2}{+7} & \numsci{5.9}{+8} & \numsci{1.1}{+10} \\  
			$N_\textrm{mem}$ [\SI{}{\giga\byte}]& \numsci{5.5}{-7} & \numsci{1.2}{-5} & \numsci{2.6}{-4} & \numsci{5.3}{-3} & \numsci{1.1}{-2} & \numsci{2.0}{+0} & \numsci{3.7}{+1} & \numsci{6.8}{+2} \\  
		\end{tabular}
		\caption{Number of states $\nstates$, maximum number of non-zero elements of the Hubbard-Hamiltonian $N_{\textrm{nz}}$, and estimated memory consumption $N_\textrm{mem}$ in gigabyte for a system with $\nsites$ sites and $\nup=\ndown=\nsites/2$. Note that $N_\textrm{nz}$ and hence also $N_\textrm{mem}$ are just upper bounds, the real memory consumption may be much lower.}
		\label{tab:dimensions}
	\end{centering}
\end{table}

\paragraph{CSR-format}

In the light of the previous considerations, the most suitable storage format for the matrix representation $H$ of the Hamiltonian is the Compressed-Sparse-Row (CSR) format \cite{csr}.
This format stores only the non-zero elements and their positions within the matrix.
For $H \in \C^{\nstates \times \nstates}$ it consists of three arrays:
\begin{itemize}
	\item $\val \in \C^{N_\textrm{nz}}$ consists of all non-zero elements of $H$ in the order they appear in $H$ in a row-wise fashion.
	\item $\col \in \N^{N_\textrm{nz}}$ consists of the column indices of all non-zero elements in the same order as in $\val$.
	\item $\row \in \N^{\nstates + 1}$ stores where the rows in $\col$ are. Its $k$-th element refers to the position in $\col$ where the $k$-th row begins and the $(k-1)$-th ends.
\end{itemize}

If $H_{ij}$ is the $k$-th non-zero element of $H$, it can thus be accessed via $\val_{k}$, and there holds $j = \col_k$ as well as $\row_i \leq k < \row_{i+1}$.
Due to the Hamiltonian being Hermitian, its matrix representation satisfies $H^\dagger=H$ and only the upper triangular part needs to be stored explicitly, i.e.\ $H_{ij}$ for all $i \leq j$.
For all other elements there holds $H_{ji} = {H^\ast_{ij}}$.
The following example illustrates this concept:
\begin{equation}
	H = \left(
	\begin{array}{ccccc}
		\mathnode{A1}{1} & \gl{0} & 1 & \iu & \mathnode{E1}{\gl{0}}\\
		\gl{0} & \mathnode{A2}{2}  & \gl{0} & \gl{0} & \mathnode{E2}{1}\\
		\gl{1} & \gl{0} & \mathnode{A3}{\gl{0}} & \gl{0} & \mathnode{E3}{\gl{0}}\\
		\g{-\iu} & \gl{0} & \gl{0} &  \mathnode{A4}{4} & \mathnode{E4}{\gl{0}}\\
		\gl{0} & \g{1} & \gl{0} & \gl{0} & \mathnode{A5}{5}
	\end{array}
	\right),
	\qquad
	\begin{array}{rcl}
	\val &=& [\,1,\, 1,\, \phantom{1}\mathclap{\iu~},\, 2,\, 1,\, 4,\, 5\,],\\
	\col &=& [\,1,\, 3,\, 4,\, 2,\, 5,\, 4,\, 5\,],\\
	\row &=& [\,1,\, 4,\, 6,\, 6,\, 7,\, 8\,].
	\end{array}
\end{equation}
\begin{tikzpicture}[overlay]%
	\draw [>=stealth, ->] (A1.west)++(-0.7em,0) to (A1.west);%
	\draw [>=stealth, ->] (E1.east) to[out=-45,in=135,looseness=1.0] (A2.west);%
	\draw [>=stealth, ->] (E2.east) to[out=-45,in=135,looseness=1.2] (A3.west);%
	\draw [>=stealth, ->] (E3.east) to[out=-45,in=135,looseness=1.5] (A4.west);%
	\draw [>=stealth, ->] (E4.east) to[out=-45,in=135,looseness=2.5] (A5.west);%
\end{tikzpicture}%
Only the upper triangular part of $H$ is considered and the non-zero elements are stored.
Note that due to row 3 having no non-zero element above the diagonal, there holds $\row_3 = \row_4 = 6$.

A comparison of memory requirement between naive and sparse (CSR) implementation of the Hamiltonian matrix for the worst case (half-filling) is shown in Table~\ref{tab:dimensions}.
Due to the small number of elements that have to be stored, the addition of matrices with the same sparsity structure can be carried out efficiently by adding the $\val$ arrays of both matrices.
Furthermore, because of the row-wise storage of the matrix, the CSR format is predestined for matrix-vector multiplication. Both of which can be done in $\mathcal{O}(N_\textrm{nz})$ operations.
The drawbacks of this format, however, lie in element-access for which a linear search of the $\col$ array must be carried out, and in changing the sparsity structure (i.e. set a former zero element to a value other than zero), in which case all three arrays must be altered and possibly reallocated. This is avoided in our implementation.

\subsection{Time dependent Hamiltonian}\label{sec:timedependent-hamiltonian}

We assume the hopping amplitudes to be time dependent in the following way:
We consider a Hermitian matrix $v^\textrm{Re} \in \C^{\nsites \times \nsites}$ and an Anti-Hermitian matrix $v^\textrm{Im} \in \C^{\nsites \times \nsites}$, as well as a phase factor $f(t) \in \C$ which vanishes for large times, i.e.\ $|f(t)| = 1$ and $f(t) \to 1$ as $t \to \infty$.
For each hopping pair $(i,j)$, we can decide if the corresponding hopping amplitude should explicitly depend on time or not.
Then, the time dependent hopping amplitudes read
\begin{equation}
\label{eq:def-hopping}
	v_{ij}(t) =
	\begin{cases}
		v^\textrm{Re}_{ij} \Re{f(t)} + \iu v^\textrm{Im}_{ij} \Im{f(t)}
		& \text{if hopping is time dependent,}\\
		v^\textrm{Re}_{ij} & \text{else.}
	\end{cases}
\end{equation}
Note that this definition renders the matrix $v(t)$ Hermitian, i.e.\ $v^\dagger(t) = v(t)$, and $v_{ii}(t)\equiv v_{ii} \in \R$.
The function $f(t)$ in~\eqref{eq:def-hopping} can e.g. describe the EM pulse as in Eq.~\eqref{eq:vt}.

By separating time-dependent and time-independent parts of the Hamiltonian according to~\eqref{eq:def-hopping}, the full time-dependent matrix representation can be written as
\begin{equation}\label{eq:hamimplementation}
	H(t) = H^\textrm{(stat)} + \Re{f(t)} H^\textrm{(Re)} + \iu \Im{f(t)} H^\textrm{(Im)}.
\end{equation}
Here, the matrix $H^\textrm{(stat)}$ includes all time-independent contributions to the Hamiltonian. These are the Coulomb interaction $U$ and hopping amplitudes $v_{ij}^\textrm{Re}$ if hopping between sites $i$ and $j$ is modeled as time independent.
The matrices $H^\textrm{(Re)}$ and $H^\textrm{(Im)}$ include all hopping amplitudes $v^\textrm{Re}_{ij}$ and $v^\textrm{Im}_{ij}$, respectively, which are modeled as time dependent.
Due to the function $f(t)$ converging to one at large times $t$, the Hamiltonian for such $t$ is $H(t) = H^\textrm{(stat)} +  H^\textrm{(Re)}$, which describes the system in equilibrium.

We suppose that $H^\textrm{(full)}$, $H^\textrm{(stat)}$, $H^\textrm{(Re)}$, and $H^\textrm{(Im)}$ can be described by only one pair of index arrays $\row$ and $\col$.
The assembly of this structure is shown in Alg.~\ref{alg:structure}.
{Because of the nested for-loops,} this costs $\mathcal{O}(\nstates^2)$ operations and is the only operation in our code that has quadratic complexity.
However, the assembly is done without knowledge of either $U$, or $v$, so the structure is independent of the interaction between sites and hence of the geometry.
Therefore, the structure for a specific set of $\nsites, \nup,$ and $\ndown$ only needs to be computed once (which can be done in parallel); {cf.\ the discussion in Sec.~\ref{subsec:limitations}.}

{To improve the complexity of Alg.~\ref{alg:structure} one needs to avoid the innermost for-loop, which can be done in the following way.
For every state $\ket{\psi_i}$, instead of comparing it to every other state in $\mathcal{B}$, one can simulate hopping of electrons as is done in Alg.~\ref{alg:genstates}.
Thus, one obtains all states that differ from $\ket{\psi_i}$ by exactly two entries, i.e., all states that interact with $\ket{\psi_i}$ and give a possibly non-zero contribution to the hamiltonian, resulting in an overall cost of $\mathcal{O}(N_\textrm{nz})$.
To achieve linear complexity in $N_\textrm{nz}$, however, it is crucial that finding the position in $\mathcal{B}$ of a given state can be done in constant time, e.g., by a suitable hash function as described in Sec.~\ref{subsec:nonequilibrium}.}

\begin{algorithm}[h]
	\caption{Computing the structure of the Hamiltonian}
	\label{alg:structure}
	\textbf{Input:} $\nstates, \mathcal{B}$\\
	\textbf{Output:} $\row, \col$
	\begin{algorithmic}[1]
		\STATE allocate arrays $\row \in \N^{\nstates+1}$ and $\col \in \N^{N_\textrm{nz}}$
		\STATE $k=1$
		\FOR{$i=1$ \TO $\nstates$}
			\STATE $\row_i = k$
			\FOR{$j=i$ \TO $\nstates$}
				\IF{$\ket{\psi_i}$ and $\ket{\psi_j}$ differ by 2 entries or less}
					\STATE $\col_k = j$
					\STATE $k = k+1$
				\ENDIF
			\ENDFOR
		\ENDFOR
		\STATE $\row_{\nstates+1} = k+1$
	\end{algorithmic}
\end{algorithm}

Pre-assembling the structure allows for fast assembly of the Hamiltonian for specific coefficients $U$ and $v$, which is shown in Alg.~\ref{alg:assembly}. 
Note that the first entry in each row of the sparse representation of the Hamiltonian lies on the diagonal, thus line~4 adds a diagonal contribution.
Furthermore, as noted in Sec.~\ref{sec:symmetries}, applying $\creation{i\sigma}\annihilation{j\sigma}$ to a state where hopping from $j\sigma$ to $i\sigma$ is possible results in a factor
\begin{equation}
\label{eq:sign-explanation}
    (-1)^{\delta(i,j,\sigma)}
    =
    (-1)^{|c_{i\sigma} - c_{j\sigma}|}.
\end{equation}
Thus, $\delta(i,j,\sigma) = |c_{i\sigma} - c_{j\sigma}|$ is the number of electrons that lie between the $i\sigma$ and $j\sigma$ entry and can be computed by a simple for-loop.
This explains the signs in lines~8~and~10.

It is apparent that the cost of Alg.~\ref{alg:assembly} is $\mathcal{O}(N_\textrm{nz})$ and that the memory consumption of the Hamiltonian is proportional to $N_\textrm{nz}$.
Upper bounds for the memory consumption for certain parameters $\nsites$, $\nup$, and $\ndown$ are shown in Table~\ref{tab:dimensions}.
To further reduce the memory consumption and computational effort for the time-stepping, one can carry out the assignments in lines~6--12 only if at least one of the contributions that would be assigned is non-vanishing.
Afterwards, the structure can be updated to only account for the actual non-vanishing elements of the Hamiltonian.

\begin{algorithm}[h]
	\caption{Assembling all parts of the Hamiltonian}
	\label{alg:assembly}
	\textbf{Input:} $\nstates, \mathcal{B}, U, v$, precomputed $\row, \col$\\
	\textbf{Output:} $H^\textrm{(stat)}, H^\textrm{(Re)}, H^\textrm{(Im)}$
	\begin{algorithmic}[1]
		\STATE allocate arrays $H^\textrm{(stat)}, H^\textrm{(Re)}, H^\textrm{(Im)} = 0 \in \C^{N_\textrm{nz}}$
		\STATE $k=1$
		\FOR{$i=1$ \TO $\nstates$}
			\STATE $H^\textrm{(stat)}_{k} = U \sum_{i=1}^{\nsites} \bra{\psi_i}\operator{d_i}\ket{\psi_i}$
			\FOR{$j=\row_i$ \TO $\row_{i+1}$}
				\STATE determine sites $\alpha$ and $\beta$ between which the hopping $\ket{\psi_{\col_j}} \rightarrow \ket{\psi_i}$ happens
				\IF{hopping between sites $\alpha$ and $\beta$ is time-dependent}
					\STATE $H^\textrm{(Re)}_{k} = (-1)^{\delta(\alpha,\beta,\sigma)} v^\textrm{(Re)}_{\alpha\beta},
					\quad H^\textrm{(Im)}_{k} = (-1)^{\delta(\alpha,\beta,\sigma)} v^\textrm{(Im)}_{\alpha\beta}$
				\ELSE
					\STATE $H^\textrm{(stat)}_{k} = H^\textrm{(stat)}_{k} + (-1)^{\delta(\alpha,\beta,\sigma)} v^\textrm{(Re)}_{\alpha\beta}$
				\ENDIF
				\STATE $k=k+1$
			\ENDFOR
		\ENDFOR
	\end{algorithmic}
\end{algorithm}

\subsection{Time-stepping algorithm}\label{sec:timestepping}

Each state in $\hilbertspace_\ndown^\nup(\nsites)$ can be uniquely represented by a vector $\vec{v} \in \R^\nstates$.
Hence, we can write the ODE system resulting from the time dependent Schr\"odinger equation Eq.~\eqref{eq:schroedinger} as
\begin{equation}\label{eq:odesystem}
	\iu \tfrac{\d{}}{\d{t}}\vec{v}(t) = H(t) \vec{v}(t), ~~ \vec{v}(0) = \vec{v}_0,
\end{equation}
where $H \in \C^{\nstates \times \nstates}$ is the matrix representation of $\operator{H}$ and $\vec{v}_0$ is the vector representing the ground state of the system.
We now discuss how to solve~\eqref{eq:odesystem} numerically.

\paragraph{Ground state}

In order to obtain the initial state for the time-stepping, we consider the system described by Eq.~\eqref{eq:schroedinger} to be in thermal equilibrium.
Then, the ground state $\ket{\psi_0} \in \hilbertspace_\ndown^\nup(\nsites)$ is defined as the eigenstate corresponding to the smallest eigenvalue of $\operator{H}$:
\begin{equation}
	\operator{H}\ket{\psi_0} = E_0 \ket{\psi_0}, ~~ E_0 = \min\set{E}{E \text{ is an eigenvalue of } \operator{H}}.
\end{equation}
Numerically, we obtain a representation $(E_0, \groundstate)$ of the eigenpair $(E_0, \psi_0)$ by a variant of the so-called power iteration method (see e.g.\ \cite{numerik}).

The power iteration method iteratively computes the eigenvalue $\lambda$ of largest absolute value and the corresponding eigenvector $\vec{v}$ of a Hermitian matrix $M \in \C^{N \times N}$ by the recursive formulae
\begin{equation}
	\vec{v}^{(n)} = \frac{M \vec{v}^{(n-1)}}{\|M \vec{v}^{(n-1)}\|}, ~~ \lambda^{(n)} = (\vec{v}^{(n-1)})^\dagger M \vec{v}^{(n-1)},
\end{equation}
starting with an arbitrary vector $\vec{v}^{(0)}$ not orthogonal to the desired vector.
The iteration stops if $\lambda^{(n)}$ and $\vec{v}^{(n)}$ are sufficiently near to the real values, which is determined by an a-posteriori error estimate.
This is shown in Alg.~\ref{alg:power-iteration}.

\begin{algorithm}[h]
	\caption{Power iteration method}
	\label{alg:power-iteration}
	\textbf{Input:} Hermitian matrix $M \in \C^{N \times N}$, maximum number of iterations $N_\textrm{max}$, $\tol$\\
	\textbf{Output:} approximate eigenpair $(\lambda, \vec{v})$ of $M$
	\begin{algorithmic}[1]
		\STATE initialize $\vec{v}^{(0)}$ randomly and normalize, $\lambda^{(0)} = 0$
		\FOR{$i=1$ \TO $N_\textrm{max}$}
		\STATE $\vec{v}^{(i)} = H\vec{v}^{(i-1)}$
		\STATE $\lambda^{(i)} = (\vec{v}^{(i-1)})^\dagger \vec{v}^{(i)}$
		\STATE $\vec{v}^{(i)} = \vec{v}^{(i)} / \|\vec{v}^{(i)}\|$
		\IF{$\abs{\lambda^{(i)}-\lambda^{(i-1)}} \leq \tol \abs{\lambda^{(i-1)}}$ \AND $\|\vec{v}^{(i)}-\vec{v}^{(i-1)}\| \leq \tol \|\vec{v}^{(i-1)}\|$}
		\STATE break
		\ENDIF
		\ENDFOR
		\STATE $\lambda = \lambda^{(i)}$, $\vec{v} = \vec{v}^{(i+1)}$
	\end{algorithmic}
\end{algorithm}

Applying the power iteration to $H$ gives an approximate eigenpair $(E,\vec{v})$.
If $E$ is negative, we have already found the smallest eigenvalue of $H$ and we can set $E_0 = E$, and $\groundstate = \vec{v}$.
Else, if $E \geq 0$ and hence is the largest eigenvalue of $H$, we apply the power iteration once again to the shifted matrix $H-EI$, obtaining an approximate eigenpair $(E', \vec{v}')$.
Since $H-EI$ has only non-positive eigenvalues, $E'$ approximates its smallest eigenvalue.
Then, we set $E_0 = E+E'$ and $\groundstate = \vec{v}'$, as they approximate the smallest eigenvalue of $H$ and the corresponding eigenvector, respectively.
This is shown in Alg.~\ref{alg:iteration}.

\begin{algorithm}[h]
	\caption{Computing the ground state}
	\label{alg:iteration}
	\textbf{Input:} matrix representation $H$ of hamiltonian, $N_\textrm{max}$, $\tol$\\
	\textbf{Output:} approximate groundstate $\groundstate$ and energy $E_0$
	\begin{algorithmic}[1]
		\STATE obtain $(E,\vec{v})$ from applying Algorithm~\ref{alg:power-iteration} to $(H,N_\textrm{max},\tol)$
		\IF{$E<0$}
			\STATE $\groundstate = \vec{v}$, $E_0 = E$
			\RETURN
		\ENDIF
		\STATE obtain $(E',\vec{v}')$ from applying Algorithm~\ref{alg:power-iteration} to $(H-EI,N_\textrm{max},\tol)$
		\STATE $\groundstate = \vec{v}'$, $E_0 = E+E'$
	\end{algorithmic}
\end{algorithm}

\paragraph{Exponential midpoint rule and Krylov subspace method}

The continuous evolution solving Eq.~\eqref{eq:odesystem} is the discrete analog of Eq.~\eqref{eq:exact-solution}.
For small times $t$, it can be approximated with sufficient accuracy by a Magnus-expansion of order zero \cite{magnus}, which gives
\begin{equation}
\label{eq:magnus}
	\vec{v}(t) \approx \exp\left( -\iu \int_0^t H(\tau) \d{\tau} \right) \vec{v}_0.
\end{equation}

By approximating the integral in the exponent via the midpoint rule
\begin{equation}\label{eq:midpoint-rule}
	\int_{a}^{b} f(t) \d{t} \approx (b-a)f\left(\frac{b+a}{2}\right),
\end{equation}
the approximation in Eq.~\eqref{eq:magnus} can be further simplified.
Considering consecutive intervals of length $\tau$, for which the midpoint rule and the Magnus-expansion are sufficient approximations, yields a sequence of vectors defined by
\begin{equation}
\label{eq:time-stepping}
	\vec{v}^{(n+1)} = \exp\left( -\iu H(n\tau + \tau/2) \tau\right) \vec{v}^{(n)},
	\quad
	\vec{v}^{(0)} = \vec{v}_0,
\end{equation}
that approximate the solution at times $n\tau$: $\vec{v}^{(n)} \approx \vec{v}(n\tau)$.
Note that these approximations are of lowest order and thus the time stepping cannot, in general, be expected to surpass first-order convergence.
{The discretization in Eq.~\eqref{eq:time-stepping} is the lowest-order representative of  a family of numerical time propagation schemes called Magnus integrators \cite{ahkqt19}.
Other representatives, which allow for higher-order time stepping methods, can be obtained by using higher-order expansions in Eq.~\eqref{eq:magnus} and Eq.~\eqref{eq:midpoint-rule} (cf.~\cite{koch}).
Of particular practical interest among these are so-called commutator free exponential time-propagators (CFETs), which avoid commutator terms in the Magnus-expansion Eq.~\eqref{eq:magnus} and thus optimize the number of necessary matrix-vector multiplications; see~\cite{Alvermann2012} for a fourth-order CFET for the Schr\"odinger equation (and a comparison to conventional Runge--Kutta-type integrators) and~\cite{Alvermann2011} for the derivation of general higher-order CFETs which can be readily implemented with the data structures and methods presented in our work.}

The main difficulty in the computation of Eq.~\eqref{eq:time-stepping} is evaluating the exponential of the large anti-Hermitian sparse matrix $-\iu H\tau$.
To this end we employ a so called Krylov subspace method as described in \cite{expokit} and references therein.
For a matrix $H$ and a vector $\vec{v}$, the $m$-th Krylov subspace is defined as $\mathcal{K}_m(H,\vec{v}) = \textrm{span}\{\vec{v}, H\vec{v}, \ldots, H^{m-1}\vec{v}\}$. The space $\mathcal{K}_m(H,\vec{v})$ is thus spanned by vectors obtained by (sparse) matrix-vector multiplication only, which can be carried out efficiently in the CSR-format.
Let $V \in \C^{\nstates \times m}$ be a projection to an orthonormal basis of $\mathcal{K}_m(H,\vec{v})$.
Then, by projection, we can approximate $H$ by a lower dimensional matrix $h \in \C^{m\times m}$:
\begin{equation}
	H \approx V h {V}^\dagger.
\end{equation}
Hermiticity of $H$ implies hermiticity of $h$ and basic orthogonality properties of a Krylov space (c.f. \cite{expokit}) cause $h$ to be Hessenberg, i.e. $h_{ij} = 0$ for $ i>j+1$.
Together, we can infer $h$ to be tridiagonal, hence the orthonormal basis $V$ as well as $h$ can be computed via a Lanczos-algorithm in $\mathcal{O}(m\nstates)$ operations, as is done in Alg.~\ref{alg:krylovexp}.

\begin{algorithm}[h]
	\caption{Computing a time-step by the Krylov subspace method}
	\label{alg:krylovexp}
	\textbf{Input:} $H$, $\vec{v}$, $\tau$, $N_\textrm{max}$, $\tol$\\
	\textbf{Output:} approximation to $\exp(-\iu H\tau)\vec{v}$
	\begin{algorithmic}[1]
		\STATE $\alpha = \norm{H}{}$, $\beta = \norm{\vec{v}}{}$
		\STATE $V_{:,1} = \vec{v}/\beta$, $\vec{y}^{(0)} = \vec{0}$
		\FOR{$j=1$ \TO $N_\textrm{max}$}
			\STATE $\vec{w} = H V_{:,j} - h_{j-1,j} V_{:,j-1}$
			\STATE $h_{j,j} = {V_{:,j}}^\dagger \vec{w}$
			\STATE $\vec{w} = \vec{w} - h_{j,j} V_{:,j}$
			\STATE $h_{j+1,j} = \norm{\vec{w}}{}, h_{j,j+1} = {h^\ast_{j+1,j}}$
			\STATE $V_{:,j+1} = \vec{w}/h_{j+1,j}$
			\STATE $\vec{y}^{(j)} = \exp(-\iu \tau h|_{j\times j}) (1,0,\ldots,0)^T$
			\STATE /* a-posteriori error */
			\STATE $\delta = \norm{\vec{y}^{(j)}-\vec{y}^{(j-1)}}{}/\norm{\vec{y}^{(j)}}{}$
			\IF{$\delta < 1$}
				\STATE $\varepsilon = \min(1+\norm{\vec{y}^{(j)}}{}, \tfrac{\delta}{1-\delta}\norm{\vec{y}^{(j)}}{})$
			\ELSE
				\STATE $\varepsilon = 1+\norm{\vec{y}^{(j)}}{}$
			\ENDIF
			\IF{$\varepsilon < \tol$ \OR $\abs{h_{j+1,j}} < \alpha \, \tol$}
				\STATE return $\beta V \vec{y}^{(j)}$
			\ENDIF
		\ENDFOR
	\end{algorithmic}
\end{algorithm}

For the exponentiation of a small $m$-by-$m$ tridiagonal matrix, numerically stable and efficient methods are implemented in the library Expokit \cite{expokit}, requiring $\mathcal{O}(m^3)$ operations.
Together, this leads to an approximation of the exponential in~\eqref{eq:time-stepping}:
\begin{equation*}
	\exp(-\iu H\tau)\vec{v}
	\approx \sum_n \frac{1}{n!} (-\iu V h {V}^\dagger \tau)^n\vec{v}
	= V \exp(-\iu h\tau) V^\dagger \vec{v}.
\end{equation*}

\begin{figure}
	\centering
	\includegraphics[width=0.5\linewidth]{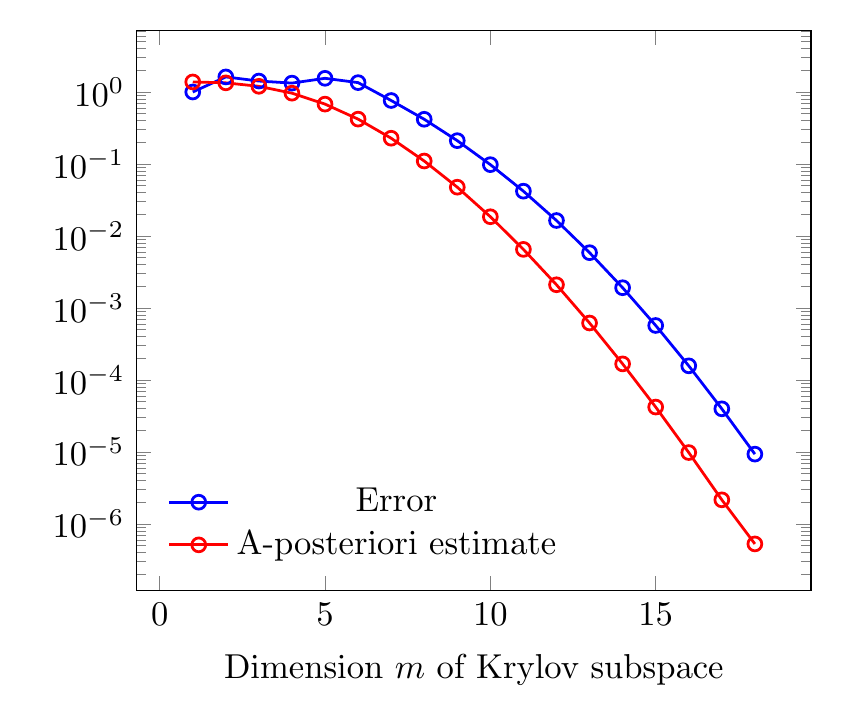}
	\caption{Comparison between actual error and a-posteriori error estimate for a diagonal problem with $\nstates = 63504$.}
	\label{fig:posteriori}
\end{figure}

The dimension $m$ of the Krylov subspace should be chosen sufficiently large to ensure small approximation errors, but small enough to limit the computational effort.
In Alg.~\ref{alg:krylovexp}, the dimension is chosen adaptively in each step via an a-posteriori error estimate, because the difficulty of computing a time-step can vary, according to $H(t)$.
We use a method suggested in \cite{posteriori}, which uses the norm of the difference of two consecutive approximations of the solution in $\mathcal{K}_{m-1}(H,\vec{v})$ and $\mathcal{K}_m(H,\vec{v})$.
The validity of this error estimate is shown in Fig.~\ref{fig:posteriori} for a problem, where the exact solution is known. Although the error estimator underestimates the error by nearly an order of magnitude, its convergence has the same rate as the error, rendering the estimate a good indicator of convergence.

Alg.~\ref{alg:krylovexp} shows one time step, summarizing this subsection.
For sake of brevity, $V_{:,j}$ denotes the $j$-th column of $V$, and $h|_{j\times j}$ denotes the restriction of $h$ to the first $j$ rows and columns.
The whole time stepping algorithm to solve Eq.~\eqref{eq:odesystem} consists of applying Alg.~6 iteratively.

\subsection{Observables}

To compute the discrete analog of the expectation value $\langle \operator{O} \rangle = \bra{\psi}\operator{O}\ket{\psi}$ of an observable $\operator{O}$ with respect to a state $\psi \in \hilbertspace_\ndown^\nup(\nsites)$, we need to discretize the action of $\operator{O}$ on $\psi$.
For the energy $\langle \operator{H}(t) \rangle$, this is already achieved by the discrete Hamiltonian $H(t)$.

For the double occupation, this can be done by computing a weight vector $\vec{w}_{\operator{d}} \in \N^\nstates$.
The $k$-th element of $\vec{w}_{\operator{d}}$ is the number of double occupations in the $k$-th state of the considered basis, i.e., $(\vec{w}_{\operator{d}})_k = \bra{\psi_k}\operator{d}\ket{\psi_k}$.
The desired expectation value can then be obtained by $\langle \operator{d} \rangle = \vec{v}^\dagger (\vec{w}_{\operator{d}} \odot \vec{v})$, where $\odot$ denotes element-wise multiplication of vectors.
For other observables like double occupation at a specific site, or electron occupation, one can compute the corresponding weight vector and proceed analogously.

\subsection{Equilibrium spectral function}

The implemented tools can also be used to compute the spectral function of the system from the Lehmann representation.
Eq.~\eqref{eq:spectralfunction} is not suitable for implementation, because of the $\delta$-distributions.
Approximating these by Lorentzians with width $\varepsilon > 0$  leads to
\begin{equation}
\label{eq:spectralplot}
	A(\nu) = \sum_{i,\sigma} \sum_{\ket{\phi}}
	\left(\frac{\varepsilon|\bra{\phi}\creation{i\sigma}\ket{\psi_0}|^2}{(\nu-E_{\ket{\phi}}+E_0)^2+\varepsilon^2}
	+\frac{\varepsilon|\bra{\phi}\annihilation{i\sigma}\ket{\psi_0}|^2}{(\nu+E_{\ket{\phi}}-E_0)^2+\varepsilon^2}\right).
\end{equation}
This approximation preserves relative values of spectral weights.

To compute $A(\nu)$ from Eq.~\eqref{eq:spectralplot}, note that, e.g.\ the creation operator $\creation{i\spinup}$ maps $\hilbertspace_\ndown^\nup(\nsites)$ to $\hilbertspace_\ndown^{\nup+1}(\nsites)$.
Then, from Eq.~\eqref{eq:directsum} it is clear that only an eigenbasis of $\hilbertspace_\ndown^{\nup+1}(\nsites)$ needs to be considered, which can be obtained by assembling the Hamiltonian in this subspace and applying an eigendecomposition.
For the evaluation of $\creation{i\spinup}\ket{\psi_0}$ the relation between states in $\hilbertspace_\ndown^\nup(\nsites)$ and $\hilbertspace_\ndown^{\nup+1}(\nsites)$ introduced by $\creation{i\spinup}$ must be known.
This relation can be found by a linear search on the states of $\hilbertspace_\ndown^{\nup+1}(\nsites)$, or, more efficiently, via hash-maps (see the following subsection).
Here, the anti-commutator relations for creation and annihilation operators have to be taken into account.

These computations can be carried out analogously for the corresponding term with annihilation operators and for the spin-down case (which can be omitted for systems with spin symmetry).

\subsection{Nonequilibrium spectral function}\label{subsec:nonequilibrium}
In order to evaluate Eq.~\eqref{eq:Glg}, the action of a creation (annihilation) operator $\creation{i\sigma}$ ($\annihilation{i\sigma}$) on a state vector $\ket{\psi}$ is implemented following the definition in Eq.~\eqref{eq:creation-annihilation}, where the sign resulting from $c_{i\sigma}$ is computed as in Eq.~\eqref{eq:sign-explanation}.


Since a general state $\ket{\psi}$ is represented as a linear combination of basis states, i.e. $\ket{\varphi_k}$
\begin{equation}
    \ket{\psi} = \sum_{k=1}^{\nstates} w(k) \ket{\varphi_k} \equiv \vec{w} \cdot \vec{\ket{\varphi}},
\end{equation}
it remains only to determine the ordering of states in the subspace with one added or one removed electron, which is not identical to the ordering of the states obtained from $\creation{i\sigma} \ket{\psi}$ or $\annihilation{i\sigma} \ket{\psi}$. Hence, after applying e.g.  $\creation{i\spinup}$ we have to find the index of the resulting state in the subspace $\hilbertspace_\ndown^{\nup+1}(\nsites)$. A simple linear search and match are very inefficient. To this aim we apply a fermionic hashing function from Ref.~\cite{hashing}  given by
\begin{equation}
\label{eq:hashing}
    I = \sum_{i = 1}^{\nsites} \colvec{p_i}{i},
\end{equation}

where $I$ is the hashing index, $p_i$ is the spin-site that the particle $i$ occupies and $\colvec{m}{n} = 0$ if $n > m$. This function provides a unique mapping of a state-vector (in its binary representation as an integer) to an integer in the range $0 \leq I < N_{\text{states}}$, which also directly corresponds to the ordering of the states. Thus, if the action of a creation (annihilation) operator on a given state-vector is non-zero, the corresponding hashing index will be calculated in order for it to be correctly assigned. 

The Fourier transform in Eq.~\eqref{eq:A_noneq} is performed as post-processing. As in case of Eq.~\eqref{eq:spectralplot} we also use broadening to numerically represent the $\delta$-distributions occurring for a finite system. This is achieved by modifying the Fourier transform in Eq.~\eqref{eq:A_noneq} by adding the factor $e^{-\varepsilon t_{rel}}$: 

\begin{equation}
    A^{\stackrel{<}{>}}(\nu,t) =  \frac{1}{\pi} \text{Im} \int_0^{T_{max}} e^{-\varepsilon t_{\text{rel}}} e^{\iu \nu t_{\text{rel}}} G^{\stackrel{<}{>}}(t,t') \d{t_{\text{rel}}},
\label{eq:A_noneq_broad}
\end{equation}
where we also limit the maximal time interval taken for the integration with sufficiently large $T_{max}$.

\subsection{Numerical cost and limitations}\label{subsec:limitations}

The method presented in this paper can handle computations of up to $14$ sites. For more than $14$ sites, some issues must be resolved.
First, indexing with a 32-bit integer format is no longer possible.
Possible remedies include using long integer indexing or splitting the arrays.
Furthermore, computation time may be an issue. The computation necessary for obtaining double occupations for a $14$-site chain (presented in  Fig.~\ref{fig:timestepping}) took about seven hours on a single node on the VSC-3 computer cluster, which is equipped with a $16$-core Intel Xeon processor and \SI{256}{\giga\byte} of RAM. Multi-node computing may accelerate simulations, e.g. by outsourcing computations of expectation values.

As seen in Sec.~\ref{sec:hamiltonian}, the time needed for the time-stepping rises with $\mathcal{O}(\nstates \ln^2(\nstates))$.
Alg.~\ref{alg:structure}, however, requires $\mathcal{O}(\nstates^2)$ operations.
{For small numbers $\nsites$, this makes up only a small amount of total runtime, since it only needs to be run once, compared to repeated simulations on the same geometry (e.g.\ parameters studies).
For a large number of sites, however, computing the structure of the hamiltonian may become the bottleneck (for $14$ sites and half-filling, it takes about one day to compute on 16 cores in parallel).
To overcome this issue, one may use the alternative algorithm described in Sec.~\ref{sec:timedependent-hamiltonian}, which has almost linear complexity.}

Also virtual memory can be an issue for more than $14$ sites, as can be seen from Table~\ref{tab:dimensions}.
For $16$ sites, the upper bound for the memory needed to store the Hamiltonian is $\SI{680}{\giga \byte}$. However, for typical geometries, many elements of $v$ are zero so that the effective memory consumption is about one fourth of the upper bound. 

\begin{figure}
	\centering
	\includegraphics[width=0.65\linewidth]{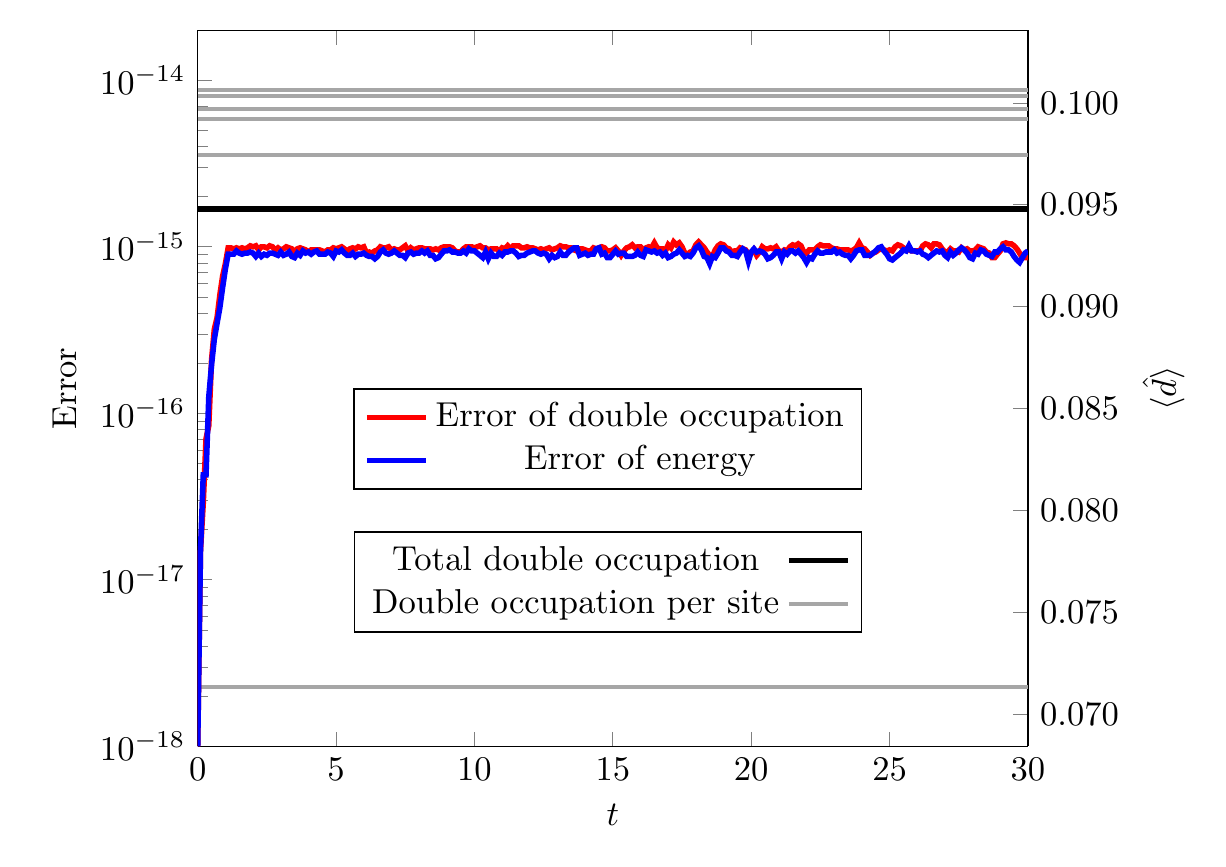}
	\caption{Errors (on the left scale) and expectation values of double occupation operators (on the right scale) for the time evolution of the ground state of a $12$-site time independent chain with $U=4$, $v_{ij}=1$ for NN sites $i$ and $j$, and half-filling. The gray lines correspond to double occupation at sites, the black line marks their mean value. Note that there are only $6$ light gray lines, because the chain geometry is symmetric with respect to its center (we use open boundary conditions). The farther away a site is from the center of the chain, the lower is its double occupation in the ground state.}
	\label{fig:stationary}
\end{figure}

\subsection{Benchmarking}

For certain values of the parameters $U$ and $v_{ij}$ the eigenvalues of the Hamiltonian from Eq.~\eqref{eq:hubbardhamiltonian} can be computed analytically.
They were compared to the numerically obtained values, and agreement within machine precision was found.
Also the eigenvalues of Hamiltonians that emanate from different geometries, for which the physics should be the same, were compared.
Again, no significant deviation was found.

For the time-stepping algorithm we tested if stationary systems are described correctly. Fig.~\ref{fig:stationary} shows double occupation and its error as well as the error of calculating the energy (difference between expected and obtained values) as a function of time for a time independent Hamiltionian. The time evolution was started from a ground state. From Fig.~\ref{fig:stationary} we can infer that the ground state is indeed an eigenstate and that the time  evolution of such a state can be correctly integrated by our algorithm.
The computation of the Green's function was benchmarked against an analytical computation for a two-site system.

\section{Results}\label{ch:results}

\begin{figure}
	\centering
	\includegraphics[width=1.0\linewidth]{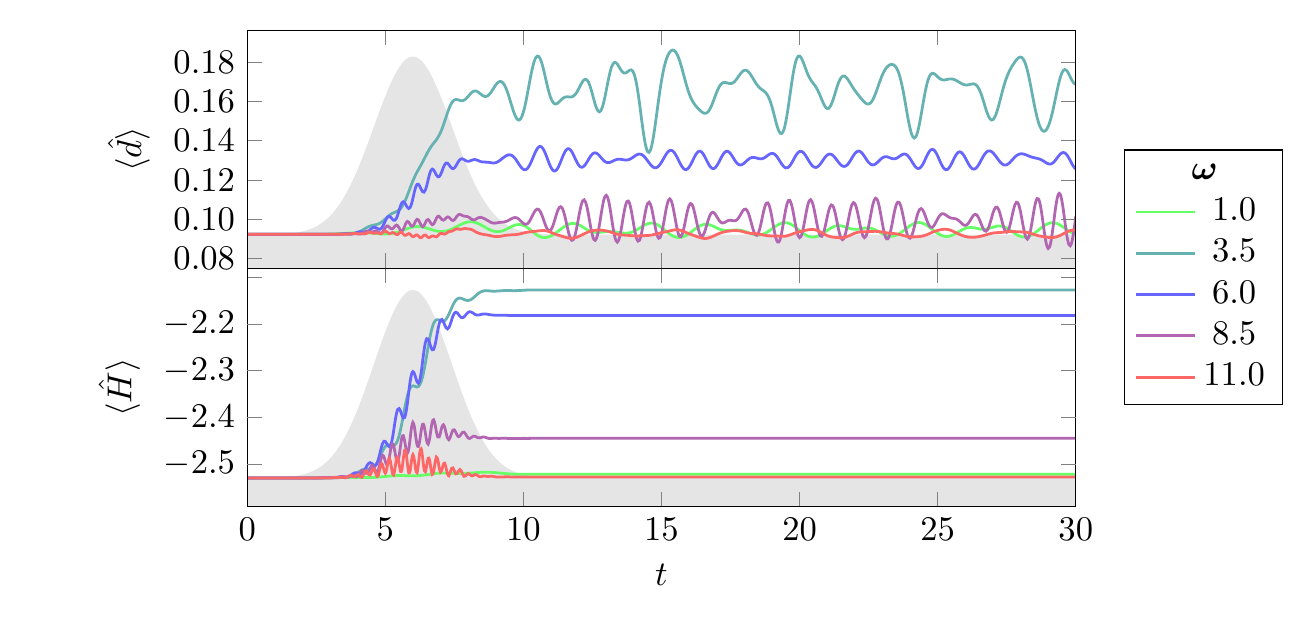}
	\caption{Average double occupation and energy per site for a half-filled $8$-site chain, $U=4$, and different pulse frequencies $\omega$. The light gray area represents the envelope of the light pulse with $t_p=6$, $\sigma=2$ and $a=0.8$.}
	\label{fig:frequencies}
\end{figure}

\begin{figure}
	\centering
	\includegraphics[width=0.8\linewidth]{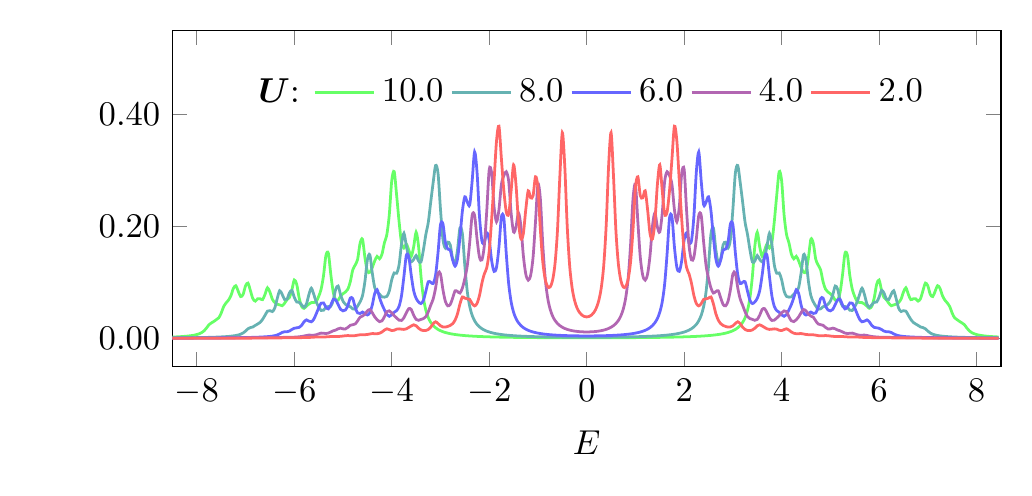}
	\caption{Average local spectral function for an $8$-site chain with half-filling for different Coulomb interaction $U$ ($\varepsilon=0.1$).}
	\label{fig:spectrum}
\end{figure}

In the following, we present results obtained for chain and box geometries with nearest neighbour (NN) hopping and open boundary conditions (OBC). We always start the time evolution from the ground state of the system and choose it to be a Mott-insulator with $n_{\spinup}=n_{\spindown}=\nsites/2$. The time step in the time-stepping algorithm is set to $\tau=0.005$, with the unit of time being 1/energy. The unit of energy (as already introduced in Sec.~\ref{ch:model}) is the absolute value of the NN hopping $|v_{ij}|=1$.

\subsection{Time evolution of double occupation}

\begin{figure}
	\centering
	\includegraphics[width=0.99\linewidth]{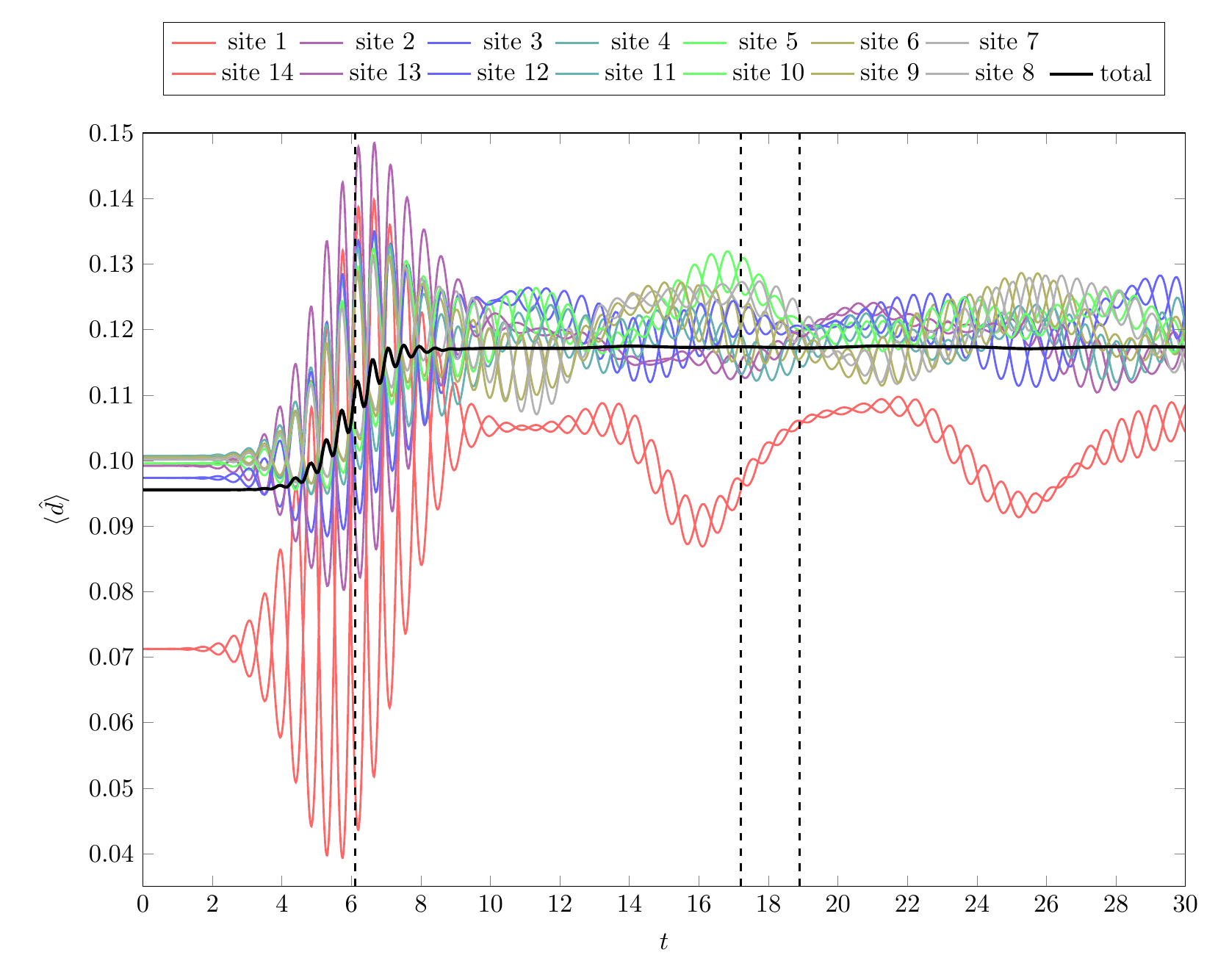}
	\caption{Time evolution of double occupation for a half-filled $14$-site chain with $U=3.5$ and pulse parameters $\omega=\tfrac{7}{4}\pi$, $t_p=6$, $\sigma=2$ and $a=0.8$. Here the colored lines represent the double occupation of the separate sites, the black line represents the average value. Sites that have the same distance from the center of the chain have the same color. The vertical dashed lines represent times, at which snapshots are shown in Fig.~\ref{fig:snapshots}.}
	\label{fig:timestepping}
\end{figure}

The effect of the light pulse described by $f(t)$ given in Eq.~\eqref{eq:vt} on the electronic system mainly depends on the relation between the pulse frequency $\omega$ and the size of the gap. In Fig.~\ref{fig:frequencies} we show the time evolution of average double occupation and energy per site for an $8$-site chain with $U=4$ for different pulse frequencies. The size of the gap for $U=4$ is approximately equal to $2$ (it can be seen in Fig.~\ref{fig:spectrum}, where we show the equilibrium spectral functions obtained from Lehmann representation for the same chain and different values of $U$). For frequencies that are significantly lower or higher than $\omega=2$ the electrons cannot be excited across the Mott gap and thus the system cannot absorb energy. Almost no electron-hole pairs are generated and the double occupation and energy stay the same after the pulse.  For the duration of the pulse only, energy is absorbed for $\omega=11$ (there is still some spectral weight in the tails of the Hubbard bands, see Fig.~\ref{fig:spectrum}), but this energy is returned to the pulse (similar deexcitation effects are described in detail within the Boltzmann equation approach in Ref.~\cite{quantum_boltzmann}).  

For frequencies $3.5-8.5$ we observe an increase of double occupation during the pulse and the increase is the strongest for $\omega=3.5$ which approximately matches the distance between the centra of the Hubbard bands. The energy is absorbed and transformed into potential energy by creating doubly occupied sites (electrons in the upper Hubbard band and holes in the lower Hubbard band).  For  $\omega=8.5$ only the lowest energy electrons can be excited to the range of the upper Hubbard band, where they occupy the high-energy part.
Thus, only a few electrons are excited but with high energies.
This causes the double occupation to barely rise, whereas the energy rises by a moderate amount.


\begin{figure}
	\centering
	\includegraphics[width=0.5\linewidth]{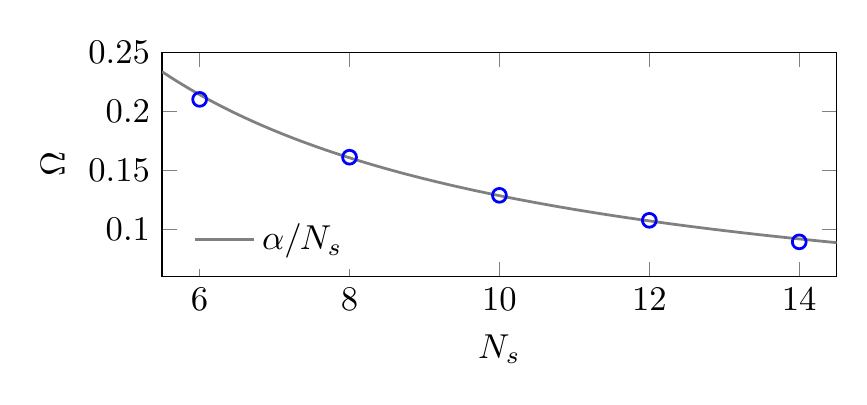}
	\caption{The dependence of the frequency $\Omega$ of the double occupation oscillations on the length of the chain $\nsites$. The gray line is an $\alpha/\nsites$ fit. }
	\label{fig:omega_low}
\end{figure}
\begin{figure}
	\centering
	\includegraphics[width=0.9\linewidth]{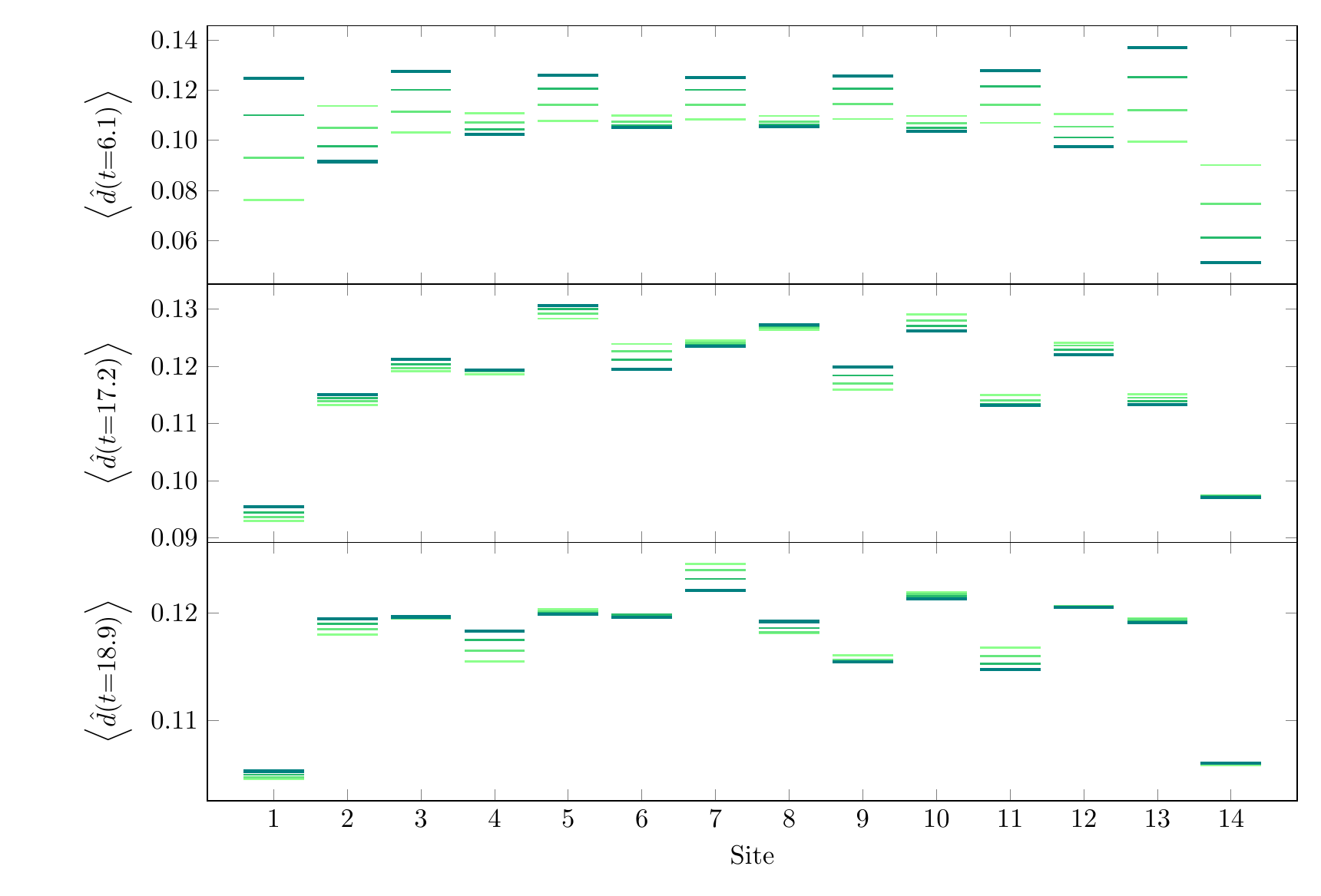}
	\caption{Snapshots of the time evolution depicted in Fig.~\ref{fig:timestepping} at different times. The darkest lines mark the value of the double occupation at the specified time $t=6.1,\;17.2\;,18.9$, the other lines are the values of the three previous time steps with $\Delta t=0.05$, increasing in saturation and darkness with time, creating the effect of the values leaving a trace. The first snapshot is taken at a phase of steepest ascend of double occupation during the pulse. The second and third snapshots are taken at a local maximum and minimum, respectively, of the total double occupation after the pulse.}
	\label{fig:snapshots}
\end{figure}

We see that, as a function of time, the double occupation oscillates with two different frequencies. This is even more visible when we look at the site resolved double occupation as presented in Fig.~\ref{fig:timestepping} for a $14$-site chain.
The high frequency oscillation is equal to the light pulse frequency $\omega$ and is typically compensated by another site where the oscillation is in opposite phase.
The lower frequency ($\Omega$) is found to be inversely proportional to the length of the chain $\nsites$ (see Fig.~\ref{fig:omega_low}). It can be viewed to originate from doublon and holon movements through the chain, leaving the overall number of doublons nearly constant. The site-averaged double occupation is almost constant in time after the pulse. We see only a slight oscillation, almost not visible in  Fig.~\ref{fig:timestepping}.

In Fig.~\ref{fig:snapshots} we show the values of double occupation along the chain for three different times: $t=6.1$ during the steep rise of double occupation during the pulse, $t=17.2$ after the pulse at a local maximum of total double occupation and $t=18.9$, at a local minimum. Initially, at $t=6.1$, an alternating pattern is visible with double occupation rising on every second site. The contribution from states, where electrons 'jump to the left' creating a doublon and leaving a hole behind is bigger than from states where electrons leave the doubly occupied sites -- in the ground state the double occupation is small. The rightmost site is different in this respect, since with the OBC there is no site to the right from which an electron could hop. At later times this alternating pattern is replaced by a longer range oscillation in space -- corresponding to doublon and holon movements along the chain. The boundary sites remain different with significantly lower double occupation due to the OBC.  

\begin{figure}
\centering
\includegraphics[width=\linewidth]{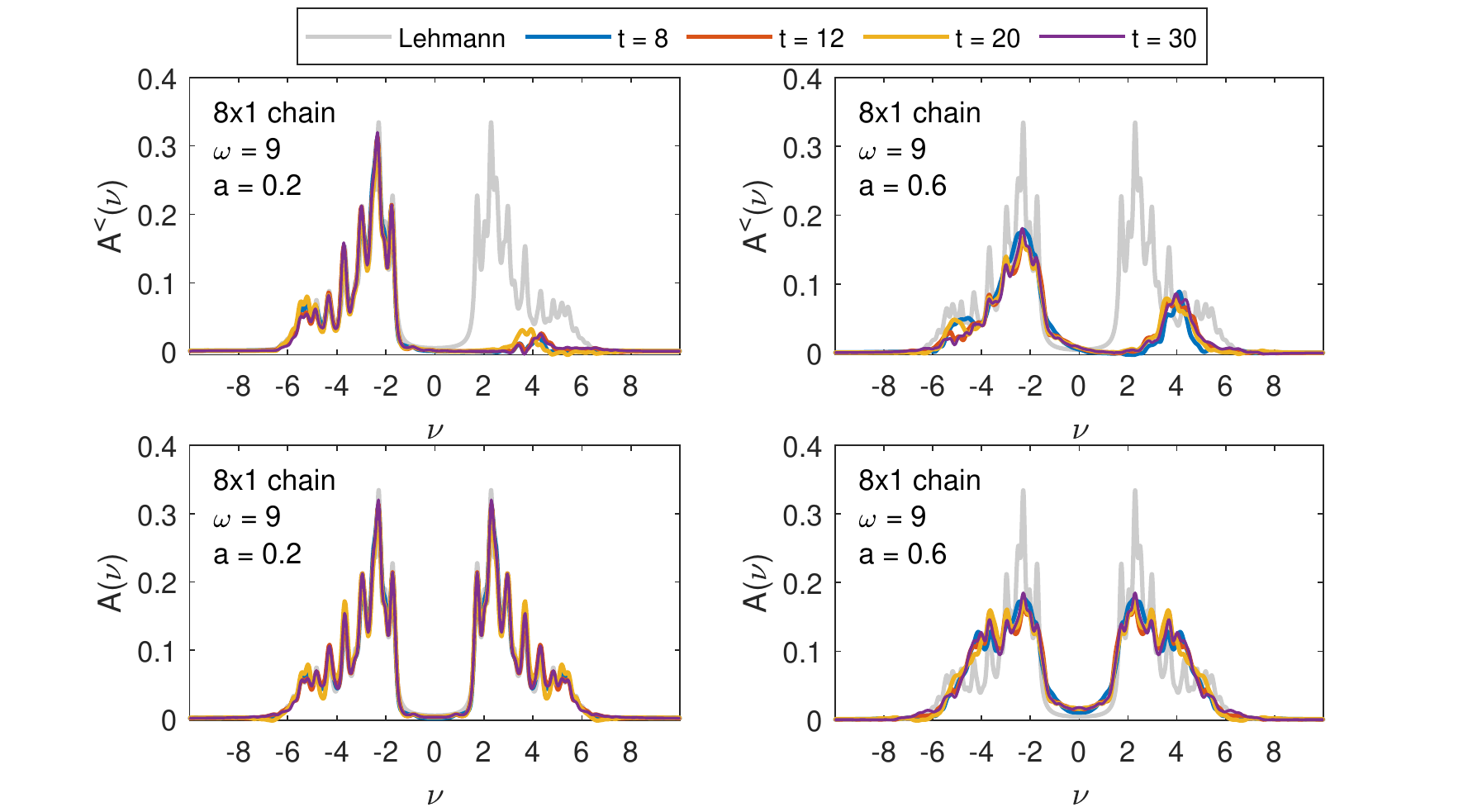}
\caption{Local, site averaged $A^{<}(\nu,t)$ (upper row) and  $A(\nu,t)$ (lower row) for different times $t = 8,\; 12,\; 20,\; 30$ and the equilibrium spectral function from Lehmann representation (grey) for an $8$-site chain with $U=6$, for pulse frequency $\omega=9$ and two different strengths of the EM field: $a=0.2$ (left column) and $a=0.6$ (right column). The Gaussian envelope of the pulse is centered at $t_p = 8$ with width $\sigma = 2$.}
\label{fig:spectral_functions_omega9_8x1}
\end{figure}

\subsection{Nonequilibrium spectral function}

\begin{figure}
\centering
\includegraphics[width=\linewidth]{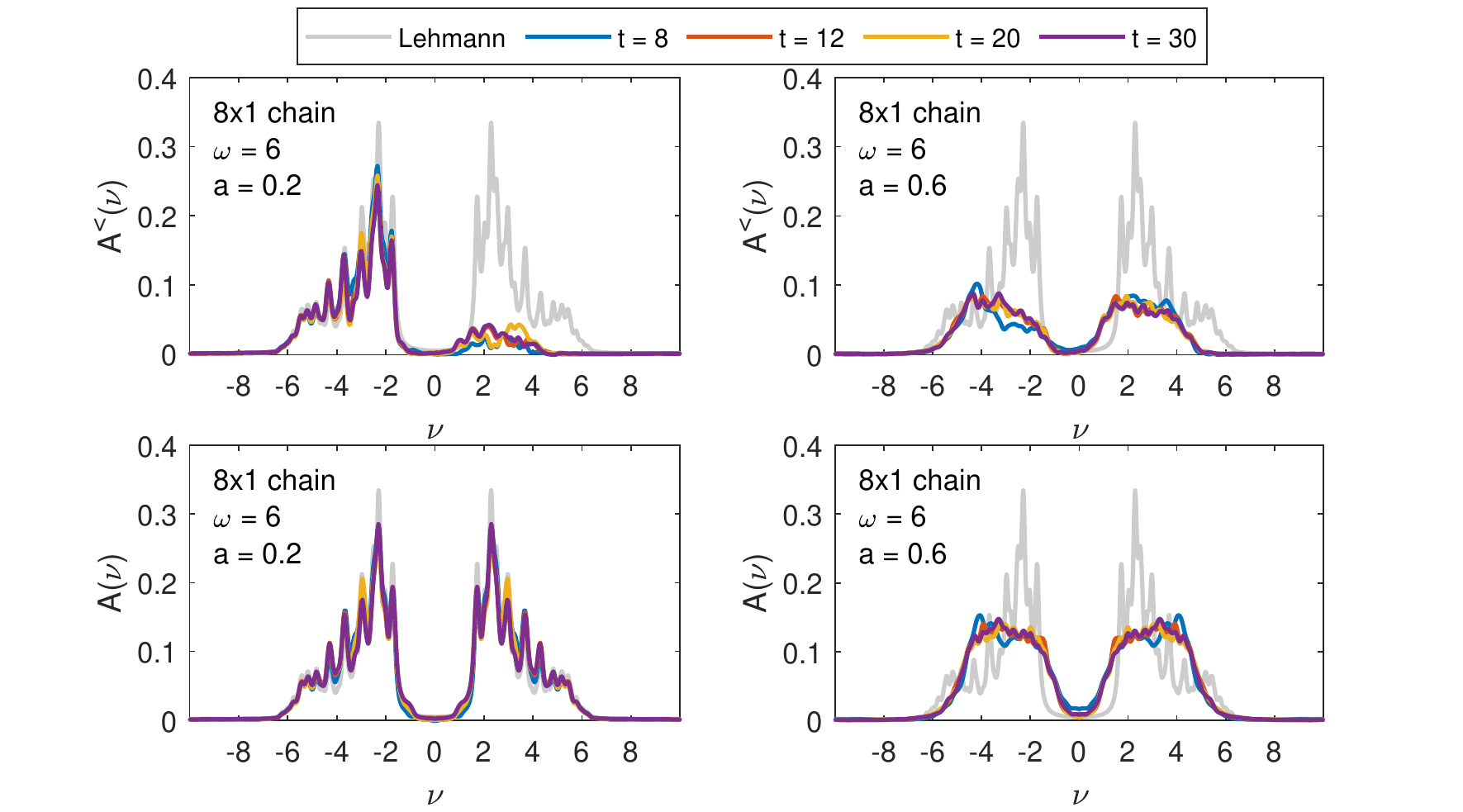}
\caption{The same as in Fig.~\ref{fig:spectral_functions_omega9_8x1} but for the pulse frequency $\omega=6$.}
\label{fig:spectral_functions_omega6_8x1}
\end{figure}

\begin{figure}
\centering
\includegraphics[width=\linewidth]{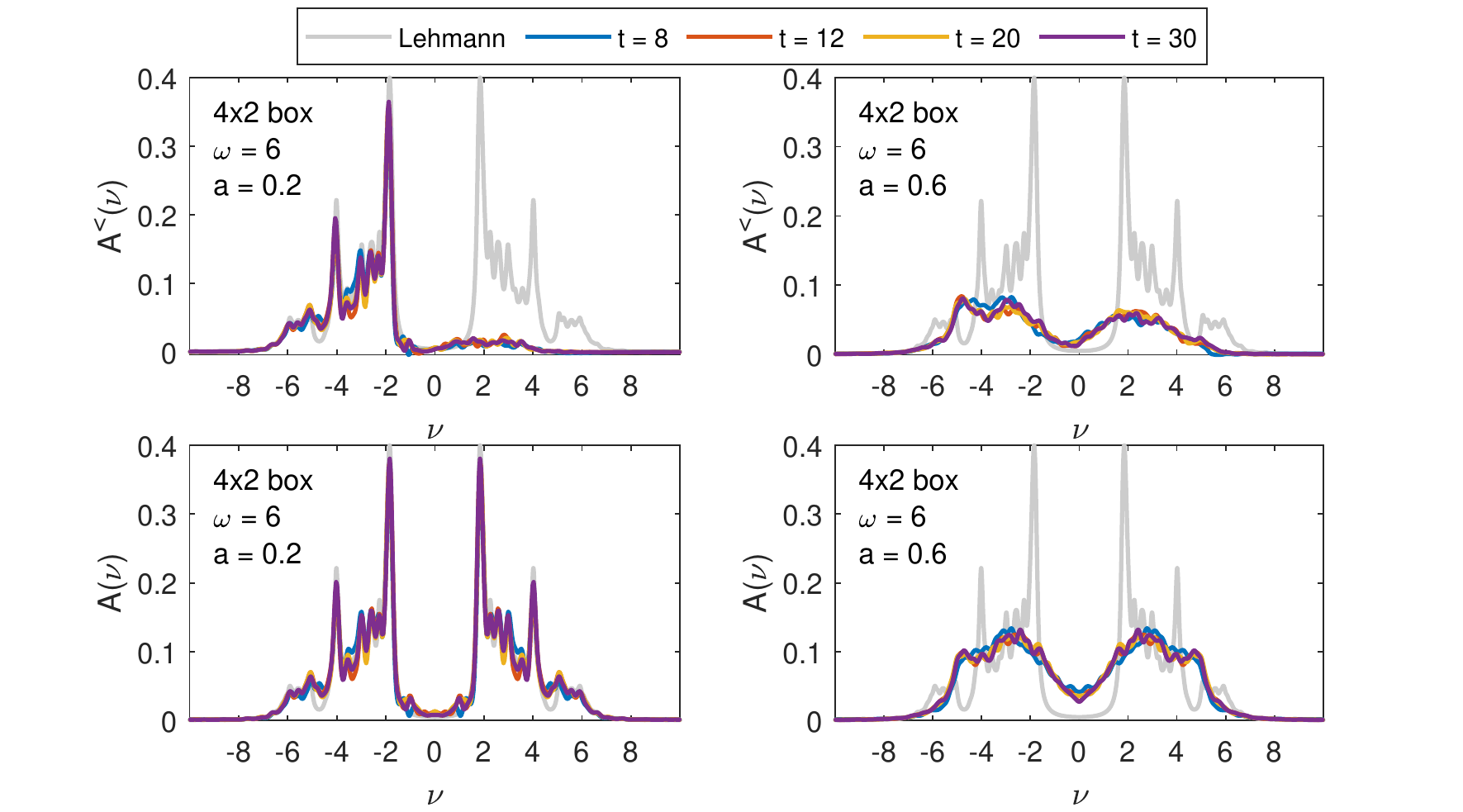}
\caption{The same as in Fig.~\ref{fig:spectral_functions_omega6_8x1}, but for the $4\times2$ box geometry.}
\label{fig:spectral_functions_omega6_4x2}
\end{figure}

The imaginary parts of the lesser Green's function $A^<(\nu,t)$ and spectral functions $A(\nu.t)$ shown in Figs.~\ref{fig:spectral_functions_omega9_8x1}--\ref{fig:spectral_functions_omega6_4x2} are calculated from Eq.~\eqref{eq:A_noneq_broad} with the broadening $\varepsilon=0.1$ and $T_{max}\approx80$. They are all local and site-averaged. Additionally we show the equilibrium spectral function in the ground state (which is our initial state at $t=0$) calculated from Lehmann representation~\eqref{eq:spectralplot} with the same broadening $\varepsilon=0.1$.

In Fig.~\ref{fig:spectral_functions_omega9_8x1} we show  $A^<(\nu,t)$ (upper row) and  $A(\nu.t)$ (lower row)  for an $8$-site chain with $U=6$ and the pulse frequency $\omega=9$ at different times during ($t=8$) and after the pulse. At $t=0$ we start from the ground state where $A^{<}$ does not have any spectral weight above $\nu=0$. For smaller pulse strength $a=0.2$ (left plots)  we see only few photo-induced excitations into the upper Hubbard band in $A^{<}$ and the overall spectrum remains almost unchanged. With increasing the pulse strength to $a=0.6$ we see stronger redistribution of the spectral weight. This effect, known as photo-doping of the Mott-insulator, has already been found in the Hubbard model with other methods -- nonequilibrium DMFT~\cite{photo_doping,photo_doping_aw} and quantum Boltzmann equation~\cite{quantum_boltzmann}. At later times, there is also an additional spectral weight shift inside both lower and  upper Hubbard bands, which corresponds to the first step of thermalization~\cite{quantum_boltzmann}. In the corresponding full spectral functions $A(\nu,t)$ we additionally see that the spectral weight shifts into the Mott-gap causing a gap reduction (photo-melting). Such gap filling is also seen in the  nonequilibrium DMFT study~\cite{photo_doping,impact}, but is missed by the quantum Boltzmann approach~\cite{quantum_boltzmann}. Both effects have also been reported in~Refs.~\cite{Okamoto_2019, Wang2017}.

 The gap filling is stronger in case we choose a smaller pulse frequency $\omega=6$ that connects points with more spectral weight than $\omega=9$ (as already noted in the discussion of Fig.~\ref{fig:frequencies}, more energy can then be pumped into the system at the same pulse intensity). The spectral function and $A^<$ of the same $8$-site chain but with $\omega=6$ are shown in~Fig.~\ref{fig:spectral_functions_omega6_8x1}. Already for $a=0.2$ there is a significant amount of spectral weight in the upper Hubbard band (upper left plot of~Fig.~\ref{fig:spectral_functions_omega6_8x1}) and we also see a slight gap filling. The shift of spectral weight into the gap already for small pulse intensity $a=0.2$ is even more pronounced for the $4\times2$ box geometry (see Fig.~\ref{fig:spectral_functions_omega6_4x2}).

\begin{figure}
\centering
\includegraphics[width=\linewidth]{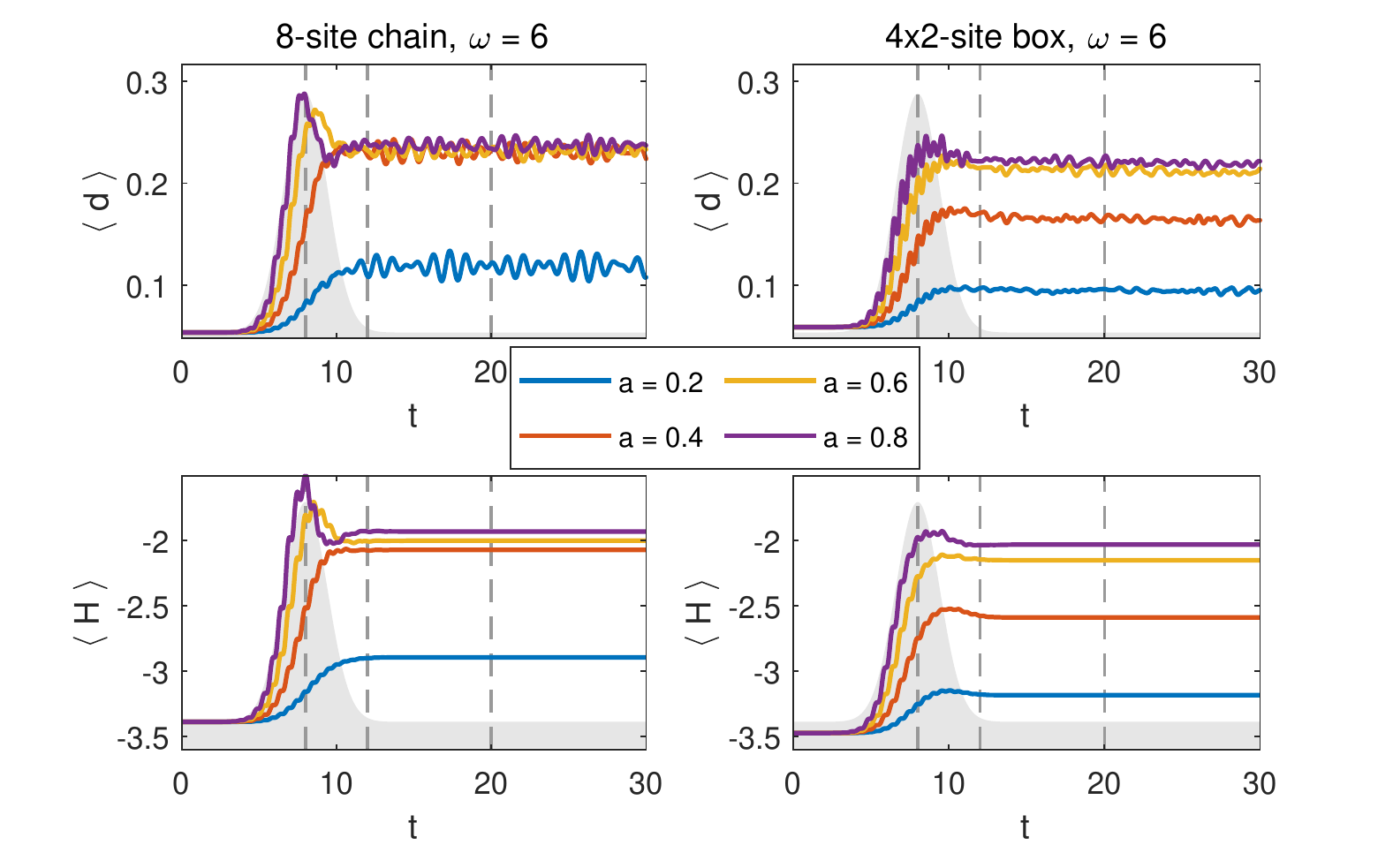}
\caption{Average double occupation (upper plots) and energy (lower plots) per site for different pulse intensities $a$ for the two geometries studied: $8\times1$ chain (left) and $4\times 2$ box (right) and $\omega=6$. Other parameters as in Figs.~\ref{fig:spectral_functions_omega6_8x1}-\ref{fig:spectral_functions_omega6_4x2}. }
\label{fig:docc_u6}
\end{figure}

 For both $8\times 1$ chain and $4\times 2$ box systems, increasing the pulse strength to $a=0.6$ causes a significant redistribution of spectral weight. Although both systems absorb approximately the same amount of energy for $a=0.6$ (cf.~Fig.~\ref{fig:docc_u6}, where we show double occupation and energy for different pulse strengths as a function of time), the gap filling is much stronger for the box geometry. In both geometries there is more spectral weight in the gap at $t=8$, i.e. during the pulse, than at later times. The systems initially absorb more energy (particularly the $8$-site chain), but it cannot be stored and is returned to the pulse (cf. Fig.~\ref{fig:docc_u6} and Ref.~\cite{quantum_boltzmann}). From Fig.~\ref{fig:docc_u6} we also learn that further increasing the pulse strength $a$ does not lead to further increase of double occupation at later times. The initial increase of energy and double occupation grows with increasing pulse strength, in case of the chain even above the maximum equilibrium value of $\langle d \rangle=0.25$, but at later times the spectral weight is redistributed and double occupation is reduced. When we look at the maximal values of energy and double occupation at a later time after the pulse (e.g. $t=20$) the chain and box geometries do not differ significantly. The chain, however, can initially absorb more energy and the rise of double occupation is bigger. This is likely related to the specific properties of the spectra, i.e. the distribution of the available states on the $\nu$-axis (the bandwidth is similar in both geometries). For the same reason, the increase of double occupation and energy at $a=0.4$ is also different for the two systems.


\section{Summary and outlook}\label{ch:outlook}

We have presented a simple implementation scheme for solving the time dependent Schr\"odinger equation for systems described by the Hubbard Hamiltonian with time dependent hoppings. As example application we show a detailed time dependence of double occupation after applying a light pulse for a $14$-site chain with open boundary conditions. We further study the photo-induced doping and gap filling of a Mott-insulator and find similar behavior for $8$-site clusters with chain and box geometry with open boundary conditions. The $8$-site clusters are certainly too small to identify differences that could come from  dimensionality (1d vs 2d) but it is an interesting future question if the chain/box small differences shown here, deepen and become more characteristic for larger systems. 

The algorithms presented here are flexible and allow for arbitrary geometry, open and periodic boundary conditions, as well as for calculating any correlation function that can be built from creation and annihilation operators. Larger cluster sizes can become possible if one can avoid storing explicitly the matrix elements of the Hamiltonian and generate them during computation. The presented implementation allows for this change, since only matrix-vector multiplications are needed. These can be replaced by operators that directly change the vector, without storing them in the matrix form. 

\section*{Acknowledgements}

The authors thank K. Held for encouragement and fruitful discussions and C. Watzenb\"ock for the help in calculating spectral functions. MI acknowledges support from the  Austrian Science Fund (FWF) through grant W 1245 and the SFB „Taming Complexity in PDE systems“ (grant F 65). PW and AK acknowledge support from the FWF through grant P 30819. Numerical computations were performed in part on the Vienna Scientific Cluster (VSC).

\begin{thebibliography}{35}

\bibitem{exp1}
D.~Fausti, R.I. Tobey, N.~Dean, S.~Kaiser, A.~Dienst, M.C. Hoffmann, S.~Pyon,
  T.~Takayama, H.~Takagi, A.~Cavalleri, Science \textbf{331}, 189 (2011)

\bibitem{exp2}
R.~Mankowsky, A.~Subedi, M.~Först, S.O. Mariager, M.~Chollet, H.T. Lemke, J.S.
  Robinson, J.M. Glownia, M.P. Minitti, A.~Frano et~al., Nature \textbf{516},
  71–73 (2014)

\bibitem{exp3}
D.~Basov, R.~Averitt, D.~Hsieh, Nat.\ Mater. \textbf{16}, 1077–1088 (2017)

\bibitem{Kaneko_eta_pairing}
T.~Kaneko, T.~Shirakawa, S.~Sorella, S.~Yunoki, Phys. Rev. Lett. \textbf{122},
  077002 (2019)

\bibitem{Fehske}
S.~Ejima, T.~Kaneko, F.~Lange, S.~Yunoki, H.~Fehske, Proceedings of the
  International Conference on Strongly Correlated Electron Systems (SCES2019)
  (2020)

\bibitem{Wang2018}
Y.~Wang, C.C. Chen, B.~Moritz, T.P. Devereaux, Phys. Rev. Lett. \textbf{120},
  246402 (2018)

\bibitem{impact2}
E.~Manousakis, Phys. Rev. B \textbf{82}, 125109 (2010)

\bibitem{impact}
P.~Werner, K.~Held, M.~Eckstein, Phys. Rev. B \textbf{90}, 235102 (2014)

\bibitem{dmft}
H.~Aoki, N.~Tsuji, M.~Eckstein, M.~Kollar, T.~Okar, P.~Werner, Rev. Mod. Phys.
  \textbf{86}, 779 (2014)

\bibitem{dca}
N.~Bittner, D.~Golež, M.~Eckstein, P.~Werner, \emph{Effects of frustration on
  the nonequilibrium dynamics of photo-excited lattice systems} (2020),
  \texttt{arXiv:2005.11722}

\bibitem{arrigoni}
D.M. Fugger, D.~Bauernfeind, M.E. Sorantin, E.~Arrigoni, Phys. Rev. B
  \textbf{101}, 165132 (2020)

\bibitem{GW}
D.~Gole\ifmmode~\check{z}\else \v{z}\fi{}, M.~Eckstein, P.~Werner, Phys. Rev. B
  \textbf{100}, 235117 (2019)

\bibitem{ED}
T.J. Park, J.C. Light, J.\ Chem.\ Phys. \textbf{85}, 5870 (1986)

\bibitem{White}
S.R. White, Phys. Rev. Lett. \textbf{69}, 2863 (1992)

\bibitem{dmrg_review1}
U.~Schollwöck, Annals of Physics \textbf{326}, 96  (2011), january 2011
  Special Issue

\bibitem{photo_doping}
M.~Eckstein, P.~Werner, Phys. Rev. Lett. \textbf{110}, 126401 (2013)

\bibitem{Eckstein2016}
M.~Eckstein, P.~Werner, Scientific Reports \textbf{6}, 21235 (2016)

\bibitem{quantum_boltzmann}
M.~Wais, M.~Eckstein, R.~Fischer, P.~Werner, M.~Battiato, K.~Held, Phys. Rev. B
  \textbf{98}, 134312 (2018)

\bibitem{photo_doping_aw}
P.~Werner, M.~Eckstein, Structural Dynamics \textbf{3}, 023603 (2016)

\bibitem{Golez2015}
D.~Gole\ifmmode~\check{z}\else \v{z}\fi{}, M.~Eckstein, P.~Werner, Phys. Rev. B
  \textbf{92}, 195123 (2015)

\bibitem{hubbard}
J.~Hubbard, Proc. Roy. Soc. A \textbf{276}, 238 (1963)

\bibitem{manybody}
G.D. Mahan, \emph{Many-particle physics} (Plenum Press, 1993)

\bibitem{Peierls}
R.~Peierls, Z. Phys. \textbf{80}, 763 (1933)

\bibitem{expokit}
R.B. Sidje, ACM Trans. Math. Softw. \textbf{24}, 130 (1998)

\bibitem{csr}
R.~Barrett, M.~Berry, T.F. Chan, J.~Demmel, J.M. Donato, J.~Dongarra,
  V.~Eijkhout, R.~Pozo, C.~Romine, H.V. der Vorst, \emph{Templates for the
  Solution of Linear Systems: Building Blocks for Iterative Methods} (Society
  for Industrial and Applied Mathematics, 1994)

\bibitem{numerik}
J.~Stoer, R.~Bulirsch, \emph{Introduction to Numerical Analysis}
  (Springer-Verlag, 1993)

\bibitem{magnus}
W.~Magnus, Comm. Pure Appl. Math. \textbf{7}, 649 (1954)

\bibitem{ahkqt19}
W.~Auzinger, H.~Hofst\"atter, O.~Koch, M.~Quell, M.~Thalhammer, ESAIM: M2AN
  \textbf{53}, 197 (2019)

\bibitem{koch}
W.~Auzinger, J.~Dubois, K.~Held, H.~Hofst\"atter, T.~Jawecki, A.~Kauch,
  O.~Koch, K.~Kropielnicka, P.~Singh, C.~Watzenb\"ock, \emph{{Efficient
  Magnus-type integrators for solar energy conversion in Hubbard models}}
  (2020), (in preparation)

\bibitem{Alvermann2012}
A.~Alvermann, H.~Fehske, P.B. Littlewood, New Journal of Physics \textbf{14},
  105008 (2012)

\bibitem{Alvermann2011}
A.~Alvermann, H.~Fehske, Journal of Computational Physics \textbf{230}, 5930
  (2011)

\bibitem{posteriori}
M.~Hochbruck, T.~Pa{\v z}ur, A.~Schulz, E.~Thawinan, C.~Wieners, ZAMM
  \textbf{95}, 237 (2014)

\bibitem{hashing}
S.~Liang, Comput.\ Phys.\ Commun. \textbf{92}, 11 (1995)

\bibitem{Okamoto_2019}
J.~Okamoto, New Journal of Physics \textbf{21}, 123040 (2019)

\bibitem{Wang2017}
Y.~Wang, M.~Claassen, B.~Moritz, T.P. Devereaux, Phys. Rev. B \textbf{96},
  235142 (2017)

\end{thebibliography}

\end{document}